# General framework for quantifying entanglement production in ultracold molecular collisions and chemical reactions

Adrien Devolder[1], Paul Brumer[1] and Timur V. Tscherbul[2]
[1]*Chemical Physics Theory Group, Department of Chemistry,
and Center for Quantum Information and Quantum Control,
University of Toronto, Toronto, Ontario, M5S 3H6, Canada*
[2]*Department of Physics, University of Nevada, Reno, NV, 89557, USA*
(Dated: January 23, 2026)

Entanglement, a defining feature of quantum mechanics, arises naturally from interactions between molecular systems. Yet the precise nature and quantification of entanglement in the products of molecular collisions and reactions remain largely unexplored. Here, we show that coupling between the external (motional) and internal degrees of freedom of the colliding molecules generates diverse forms of product-state entanglement: discrete–discrete, continuum–continuum, and hybrid discrete–continuum. We develop a general theoretical framework to quantify these entanglement forms directly from scattering S-matrix elements and identify a novel class of entangled states—multimode hybrid cat states—that exhibit multimode discrete–continuum entanglement. Although applicable at arbitrary collision energies, the formalism is illustrated in the ultracold and cold regimes for inelastic Rb + SrF and Rb + Sr$^+$ collisions, as well as the chemical reaction F + HD $\to$ HF + D, DF + H. We demonstrate that entanglement can be efficiently controlled near magnetic Feshbach resonances, paving the way for precise magnetic control of product-state entanglement generation in ultracold molecular collisions.

## I. INTRODUCTION

Recent experimental advances in the production, trapping, and coherent manipulation of ultracold molecular gases have opened up a wide array of opportunities for studying quantum phenomena in chemistry. In particular, quantum entanglement, widely recognized as a striking and counterintuitive feature of quantum physics [1–4], quite naturally emerges during molecular interactions [5–8], photodissociation [9], photoionization [10] and chemical reactions [11], redistributing across electronic, vibrational, rotational, spin, and translational degrees of freedom. Leveraging nonlocal quantum correlations between molecules embodied by entanglement could enable robust coherent control of chemical reactions [12–15], coherent many-body chemistry [16], and the generation of entangled reaction product pairs [17] for use in, e.g., quantum teleportation experiments [18] fundamental tests of quantum and gravitational physics [19, 20], and quantum interference-based coherent control [15].

The production of entangled molecular pairs has recently been experimentally realized with ultracold CaF and KRb molecules in optical tweezers [5–8]. In these experiments, molecules were held at large intermolecular separations ($R \simeq 500$ nm) and interacted only via weak long-range electric dipolar interactions [21]. This excludes the vast majority of bimolecular collisions and chemical reactions, which occur at close range, and are mediated by deep and strongly anisotropic intermolecular interactions. While these processes can also generate useful entanglement between collision/reaction products [17], it remains unclear how to classify, quantify, observe and utilize such "chemically generated" entanglement [22].

Previous work has focused on the production of two-atom entangled states in elastic [23–28] and spin-exchange collisions [29–31]. Elastic collisions can generate continuum-variable, Einstein–Podolsky–Rosen (EPR)-like entanglement between the motional degrees of freedom of structureless atoms [23, 32]. As recently shown in the pioneering treatment by Wang and Koch [28], this type of entanglement vanishes in the limit of plane-wave initial states, scales linearly with the integral scattering cross section, and is enhanced near shape resonances. Gao *et al.* suggested the possibility of observing quantum correlations between the rotational angular momenta in NO + O$_2$ collisions [33]. Bipartite and multipartite entanglement in atomic ensembles has been observed in a series of elegant experiments with interacting Bose-Einstein condensates [19, 34–38]. Recently, pairs of metastable He atoms entangled in momentum space and internal (spin) degrees of freedom, have been produced [20], which could be used for precision tests of gravitational effects on quantum states.

However, the entanglement of rotational, vibrational, fine and hyperfine degrees of freedom produced in ultracold molecular collisions and chemical reactions is yet to be explored. Compared to simple atomic collisions, molecular collisions present unique challenges and opportunities associated with their highly complex quantum dynamics mediated by an enormous number of rovibrational, fine, and hyperfine states coupled by strongly anisotropic intermolecular interactions [39]. These complicated dynamics often results in an atomic rearrangement, producing reaction products, and creating ongoing challenges to the theoretical understanding of quantum reaction dynamics [39, 40].

Here, we present a comprehensive theoretical framework for exploring, classifying, and quantifying the entanglement generated in two-body molecular collisions and chemical reactions. We find that post-collision en-tanglement can be classified into three distinct types, involving both the discrete and continuous (motional) degrees of freedom of reaction products. We quantify these entanglement types as they occur in elastic, inelastic, and reactive molecular collisions, and unravel their connection to the fundamental scattering S-matrix, which governs the collision event. We also present rigorous coupled-channel calculations on experimentally relevant systems to elucidate the entanglement-generated power of ultracold atom-molecule collisions, such as Rb + SrF, and chemical reactions, such as F + HD → HF + D.

Our results show that inelastic and reactive molecular collisions can serve as a powerful generator of entangled product pairs, and that the amount of entanglement in the product states is sensitive to the magnitudes and phases of scattering $S$-matrix elements, enabling one to probe the wavefunction of the collision complex in unprecedented detail (as in scattering state tomography [41, 42]). We further show that the product-state entanglement generated in ultracold atom-molecule collisions can be efficiently tuned by an external magnetic field. These results open up novel opportunities for generating, exploring, and utilizing quantum entanglement in ultracold molecular collisions and chemical reactions, providing a unique laboratory for studying quantum effects in molecular processes.

This manuscript is organized as follows. In Section II, we introduce the classification framework for entanglement generated in molecular collisions and chemical reactions. We distinguish three distinct types of entanglement, each treated in a dedicated section: discrete-variable–discrete-variable (DV-DV) entanglement in Section III, continuous-variable–continuous-variable (CV-CV) entanglement in Section IV, and hybrid discrete-variable–continuous-variable (DV-CV) entanglement in Section V. Finally, conclusions are presented in Section VI.

## II. CLASSIFICATION OF ENTANGLEMENT GENERATED BY MOLECULAR COLLISIONS AND CHEMICAL REACTIONS

Consider a binary collision between two quantum particles $A$ and $B$, such as molecules or atoms, initially in their internal states $\alpha_0$ and $\beta_0$ (see Fig. 1)

$$A(\alpha_0) + B(\beta_0) \to A(\alpha) + B(\beta). \quad (1)$$

As a result of the collision, the particles can either preserve their internal states (elastic collisions, $\alpha = \alpha_0$, $\beta = \beta_0$) or change them (inelastic collisions, $\alpha \neq \alpha_0$, $\beta \neq \beta_0$). Alternatively, the particles can change their chemical identity in a reactive collision. The combined system $A-B$ (a collision complex) is a bipartite quantum system, so the various types of entanglement that can arise in a collision can be generally classified according to different partitions of the full Hilbert space of the complex

$$\mathcal{H}_{\text{full}}^{inel} = \mathcal{H}_A^{int} \otimes \mathcal{H}_A^{ext} \otimes \mathcal{H}_B^{int} \otimes \mathcal{H}_B^{ext}, \quad (2)$$

where $\mathcal{H}_{A/B}^{int}$ and $\mathcal{H}_{A/B}^{ext}$ are the Hilbert spaces associated with the internal discrete-variable (DV) and external continuous-variable (CV) degrees of freedom of particles $A$ and $B$, respectively. In a two-body collision, these degrees of freedom become dynamically coupled, leading to the generation of entanglement between the internal (DV-DV) [1], external (CV-CV) [43], and between the internal and external (DV-CV) [44, 45] degrees of freedom of collision products. The three types of entanglement form the basis of our entanglement classification framework.

Entanglement of the DV-DV type is commonly encountered in standard DV-based (digital) quantum information processing [3] and is associated with the bipartition $\mathcal{H}_A^{int} \otimes \mathcal{H}_B^{int}$ with external CV degrees of freedom treated as fixed parameters. This corresponds to projecting the scattered wavefunction onto position eigenstates of particles $A$ and $B$ (via, e.g., detecting their spatial positions), and evaluating the entanglement between the internal states conditioned on this postselection. An analogous measurement strategy has been demonstrated experimentally by Shin *et al.* [19], who reported Bell-type correlations in the spin degrees of freedom of ultracold He atoms following binary collisions.

Alternatively, CV-CV entanglement [43] refers to the bipartition $\mathcal{H}_A^{ext} \otimes \mathcal{H}_B^{ext}$ with fixed internal states. Here, one projects the scattered wavefunction onto specific internal eigenstates of particles $A$ and $B$ (via, e.g., detecting these eigenstates), and quantifies the resulting entanglement between the external degrees of freedom. This form of entanglement is present even in elastic collisions and is, in that case, the only possible kind of entanglement that can be generated. The generation of CV-CV entanglement is a direct consequence of total momentum conservation, which enforces correlations between the momenta of the colliding particles. Although the individual particle momenta are unknown before measurement, a measurement of the momentum (or propagation direction) of one particle uniquely determines the corresponding momentum of the other particle as a consequence of total momentum conservation.

Finally, hybrid DV-CV entanglement [44] corresponds to the bipartition $(\mathcal{H}_A^{int} \otimes \mathcal{H}_B^{int}) \otimes (\mathcal{H}_A^{ext} \otimes \mathcal{H}_B^{ext})$. Unlike the previous cases, no postselection is required: the entanglement can be evaluated directly from the full outgoing scattering state. The hybrid entanglement arises as a consequence of the energy conservation which enforces that the values of the relative momenta are correlated to the internal states of the particles. Although the internal states of the particles and their relative momenta are unknown before measurement, a measurement of the internal states uniquely determines the value of the magnitude of the relative momenta as a consequence of energy conservation.

The classification of entanglement in reactive collisions is inherently more complex than in purely inelastic processes due to the presence of multiple possible product arrangements, such as $A + BC$, $AB + C$ or $AC + B$ in atom-diatom chemical reactions. In the context of quantum technology applications, the relevant entanglement is that between the products of a given chemical reaction; accordingly, the wavefunction is postselected on the corresponding product arrangements. For each arrangement, the Hilbert space is a tensor product of the internal and external Hilbert spaces of the product molecules, in analogy with inelastic collisions. For instance, for the arrangement $A + BC$, the Hilbert space can be expressed as $\mathcal{H}_{AB+C} = \mathcal{H}^{int}_{AB} \otimes \mathcal{H}^{ext}_{AB} \otimes \mathcal{H}^{int}_{C} \otimes \mathcal{H}^{ext}_{C}$. Within each arrangement subspace, the different types of entanglement, DV-DV, CV-CV, and hybrid DV-CV, are rigorously defined in the same manner as described above. For example, when the reaction $A + BC \to AB + C$ occurs, the CV-CV entanglement between $AB$ and $C$ is computed after a postselection on the internal states of $AB$ and $C$. Similarly, the hybrid DV-CV entanglement between the internal and external degrees of freedom of $AB$ and $C$ is evaluated.

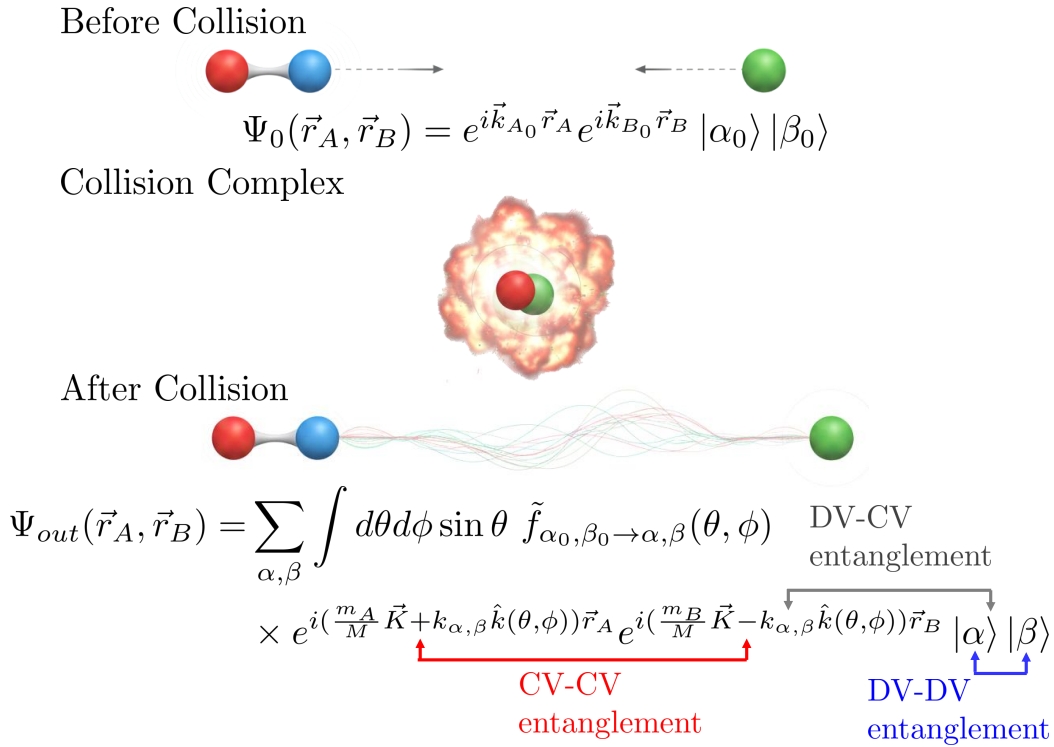

FIG. 1: Schematic representation of entanglement generation during a collision. Prior to the collision, particles A and B occupy the internal states $|\alpha_0\rangle_A$ and $|\beta_0\rangle_B$, respectively, and their translational motion is described by plane waves. At this stage, the internal and external degrees of freedom are separable, and no entanglement is present. During the collision, interactions couple the internal and translational degrees of freedom. As a result, the post-collision wavefunction becomes a coherent superposition of internal and motional states, leading to entanglement between external and internal degrees of freedom. This framework allows for the identification of three distinct classes of entanglement: discrete variable – discrete variable (DV–DV), continuous variable – continuous variable (CV–CV), and hybrid discrete variable – continuous variable (DV–CV) entanglement.

In the following sections, we will explore and quantify how the DV-DV, CV-CV, and DV-CV types of entanglement arise in elastic, inelastic, and reactive molecular collisions.

## III. DV-DV ENTANGLEMENT

### A. Formalism

Here, we consider the DV-DV entanglement that arises between the internal states of scattered particles following a two-body collision. To this end, we consider a coincidence detection scheme in which the scattered particles A and B are detected at a large interparticle separation $R_f$ in opposite direction, particle A along the direction $\hat{u}(\theta_f, \phi_f)$ and particle B along the direction $-\hat{u}(\theta_f, \phi_f)$, i.e. at angles $(\theta_f, \phi_f)$ for the particle A and at angle $(\pi-\theta_f, \phi_f+\pi)$ for the particle B. $\hat{u}(\theta_f, \phi_f)$ is the unit vector in the direction specified by the polar and azimuthal angles $(\theta_f, \phi_f)$ in the laboratory frame. Therefore, the postselected detection positions are chosen as: $\vec{r}_{A,selec} = \frac{R_f}{2}\hat{u}(\theta_f, \phi_f)$ and $\vec{r}_{B,selec} = -\frac{R_f}{2}\hat{u}(\theta_f, \phi_f)$.

Under this coincidence postselection and in the asymptotic large-$R$ regime, the scattered wavefunction reduces to (see Supplementary Material S2 for the derivation)

$$\Psi_{out}(R_f, \theta_f, \phi_f) = \frac{e^{iR_f \frac{m_A - m_B}{M} \vec{K} \hat{u}(\theta_f, \phi_f)}}{\sqrt{\frac{d\sigma(\theta_f, \phi_f)}{d\Omega}}} \sum_{\alpha, \beta} \sqrt{\frac{k_{\alpha,\beta}}{k_0}} f_{\alpha_0, \beta_0 \to \alpha, \beta}(\theta_f, \phi_f) e^{ik_{\alpha\beta} R_f} |\alpha\rangle_A |\beta\rangle_B, \quad (3)$$

where $|\alpha\rangle_A |\beta\rangle_B$ are the internal states of $A$ and $B$ in the $R \to \infty$ limit, which may correspond to the vibrational, rotational, electronic and nuclear spin states of the isolated molecules. The total differential cross section in the direction $(\theta_f, \phi_f)$ is denoted as $\frac{d\sigma(\theta_f, \phi_f)}{d\Omega}$, and $f_{\alpha_0, \beta_0 \to \alpha, \beta}(\theta_f, \phi_f)$ represents the scattering amplitude associated with the transition $|\alpha_0\rangle_A |\beta_0\rangle_B \to |\alpha\rangle_A |\beta\rangle_B$ in the direction $(\theta_f, \phi_f)$. Further, $k_0$ is the magnitude of the initial relative momentum while $k_{\alpha,\beta}$ is its value after the scattering if the state $|\alpha\rangle_A |\beta\rangle_B$ is occupied. $\vec{K}$ is the center-of-mass momentum. The quantities $m_A, m_B$, and $M = m_A + m_B$ denote the masses of particles $A$ and $B$ and the total mass, respectively. Note that $R_f, \theta_f$ and $\phi_f$ are parameters of the postselected wavefunctions and not variables.

It is worth noting that the expression in Eq. (3) is valid provided that the detection direction $(\theta_f, \phi_f)$ differs from the initial propagation direction. In the forward-scattering direction, the scattered contribution $\frac{i}{2\pi} \sqrt{k_{\alpha\beta} k_0} f_{\alpha_0, \beta_0 \to \alpha, \beta}(\theta_f, \phi_f)$ must be modified to $\delta(\Omega - \Omega_0) + \frac{i}{2\pi} \sqrt{k_{\alpha\beta} k_0} f_{\alpha_0, \beta_0 \to \alpha, \beta}(\theta_f, \phi_f)$ [46]. The non-interacting component will be seen to significantly reduce the generated entanglement. Furthermore, the presence of the delta function complicates the calculations. This could be solved by considering the initial scattering state as a narrow wavepacket localized in momentum space and regularizing the delta function [28], as done below in the calculation of the CV-DV entanglement.

From the postselected wavefunction in Eq. (3), we obtain the reduced density matrix $\rho_{\alpha,\alpha'}$ for the internal states of particle $A$ by tracing over the internal states of particle $B$ (See Supplementary Material S2 for details of the derivation):

$$\rho_{\alpha,\alpha'}(R_f, \theta_f, \phi_f) = \frac{\sum_\beta \frac{\sqrt{k_{\alpha,\beta} k_{\alpha',\beta}}}{k_0} f_{\alpha_0, \beta_0 \to \alpha, \beta}(\theta_f, \phi_f) f^*_{\alpha_0, \beta_0 \to \alpha', \beta}(\theta_f, \phi_f) e^{i(k_{\alpha\beta} - k_{\alpha'\beta}) R_f}}{\frac{d\sigma(\theta_f, \phi_f)}{d\Omega}}. \quad (4)$$

The trace induces a partial loss of coherence due to the entanglement with particle $B$. This decoherence is quantitatively captured by the entanglement entropy, which enables one quantify the amount of entanglement in a given quantum state [1]. More specifically, the entanglement entropy $S_{ent}^{int-int}$ between the internal DV degrees of freedom of collision products can be obtained as

$$S_{ent}^{int-int}(R_f, \theta_f, \phi_f) = -\left(\sum_i \lambda_i^{int-int} \log_2(\lambda_i^{int-int})\right), \quad (5)$$

where $\lambda_i^{int-int}$ are the eigenvalues of the reduced density matrix $\rho_{\alpha,\alpha'}$ in Eq. (4).

It is noteworthy that the reduced density matrix exhibits an explicit dependence on the detection distance $R_f$. Consequently, this spatial dependence can affect its eigenvalues and, hence, the associated entanglement entropy. The factor $e^{i(k_{\alpha\beta} - k_{\alpha'\beta}) R_f}$ represents a relative phase between internal states, which varies with the detection distance. This phase arises exclusively between channels corresponding to different internal energies, leading to distinct propagation phases for internal states with different energies during free evolution.

At first glance, it may appear counterintuitive that the entanglement can vary after the system has exited the interaction region. This effect originates from the postselection procedure and constitutes an example of measurement-induced entanglement [47]: when four subsystems are mutually entangled, a measurement on two of them can modify the entanglement shared by the remaining two. In the present context, the external (motional) degrees of freedom are entangled with the internal degrees of freedom, and the measurement of the external coordinates induces a corresponding change in the entanglement between the internal states. This mechanism directly accounts for the observed dependence of the reduced density matrix on $R_f$.

Analytical expressions for the DV-DV entanglement entropy can be obtained in certain specific cases. One such example is the collision between two qubits (e.g two-level atoms or molecules with a ground state $|0\rangle$ and an excited state $|1\rangle$). In this situation, Eq. (4) for the reduced density matrix becomes:

$$\rho(R_f, \theta_f, \phi_f) = \frac{1}{\frac{d\sigma}{d\Omega}} \begin{pmatrix} \Sigma_0 & C \\ C^* & \Sigma_1 \end{pmatrix}. \quad (6)$$

where:

$$\Sigma_0(\theta_f, \phi_f) \equiv \frac{d\sigma_{00}}{d\Omega} + \frac{d\sigma_{01}}{d\Omega} \quad (7)$$

$$\Sigma_1(\theta_f, \phi_f) \equiv \frac{d\sigma_{10}}{d\Omega} + \frac{d\sigma_{11}}{d\Omega}. \quad (8)$$

and

$$C(R_f, \theta_f, \phi_f) \equiv \frac{1}{k_0} \left[ f_{00} f_{10}^* \, e^{i(k_{00}-k_{10})R_f} \sqrt{k_{00}k_{10}} \right. \\ \left. + f_{01} f_{11}^* \, e^{i(k_{01}-k_{11})R_f} \sqrt{k_{01}k_{11}} \right]. \quad (9)$$

Here $f_{ij}$ denotes the scattering amplitude from the initial states to the final states $|i\rangle_A |j\rangle_B$ ($i,j=0,1$) in the postselected scattering direction $(\theta_f, \phi_f)$, while $\frac{d\sigma_{ij}}{d\Omega}$ is the corresponding differential cross section. $k_{ij}$ the magnitude of the relative momentum when the system occupies the state $|i\rangle_A |j\rangle_B$ after the scattering, and $k_0$ is the magnitude of the initial relative momentum.

Diagonalizing Eq. (6) yields the eigenvalues

$$\lambda_{1,2}^{int-int} = \frac{1 \pm \sqrt{1-4\Delta}}{2}, \quad (10)$$

$$\Delta = \frac{1}{\left(\frac{d\sigma}{d\Omega}\right)^2} \left( \mathcal{D}_1 - \mathcal{D}_2 \cos(\Phi) \right). \quad (11)$$

$$\mathcal{D}_1 = \frac{d\sigma_{11}}{d\Omega} \frac{d\sigma_{00}}{d\Omega} + \frac{d\sigma_{01}}{d\Omega} \frac{d\sigma_{10}}{d\Omega} \quad (12)$$

$$\mathcal{D}_2 = 2\sqrt{\frac{d\sigma_{00}}{d\Omega} \frac{d\sigma_{01}}{d\Omega} \frac{d\sigma_{10}}{d\Omega} \frac{d\sigma_{11}}{d\Omega}} \quad (13)$$

$$\Phi = \chi_{00} + \chi_{11} - \chi_{01} - \chi_{10} + (k_{00}+k_{11}-k_{01}-k_{10})R_f \quad (14)$$

Here, $\chi_{ij}$ are the phases of the corresponding scattering amplitudes. Equation (11) emphasizes that the discrete–discrete entanglement can explicitly depend on the relative phases of the scattering amplitudes, in contrast to the integral cross sections, which depend only on the magnitudes. *Consequently, like coherent control experiments [13], measurements of entanglement could provide a sensitive probe of the phases of the S-matrix elements, offering access to information that is otherwise inaccessible through standard cross-section measurements.*

The $R_f$-dependence occurs also in the last term of Eq. (11) through an oscillatory contribution. The visibility of this $R$-dependent modulation can be predicted from the ratio $\mathcal{D}_1/\mathcal{D}_2$ of Eqs. (12) and (13). This ratio can be interpreted as the quotient between the arithmetic and geometric means of the products $\frac{d\sigma_{11}}{d\Omega} \frac{d\sigma_{00}}{d\Omega}$ and $\frac{d\sigma_{01}}{d\Omega} \frac{d\sigma_{10}}{d\Omega}$. When this ratio is large, the oscillatory contribution is strongly suppressed and the $R$-dependence becomes negligible. The observability of the distance dependence further requires a nonvanishing momentum mismatch $\Delta k = k_{00} + k_{11} - k_{01} - k_{10}$.

Another analytically tractable scenario arises when the postselected wavefunction is expressed in a form similar to the Schmidt decomposition where each populated internal state $|\alpha_j\rangle_A$ of particle $A$ is uniquely correlated with a specific internal state $|\beta_j\rangle_B$ of particle $B$

$$\Psi_{out}(R_f, \theta_f, \phi_f) = \left(\frac{d\sigma}{d\Omega}\right)^{-1} \sum_j \sqrt{\frac{k_{\alpha_j,\beta_j}}{k_0}} f_{\alpha_j,\beta_j}(\theta_f, \phi_f) \\ \times e^{ik_{\alpha_j\beta_j}R_f} |\alpha_j\rangle_A |\beta_j\rangle_B. \quad (15)$$

In this case, the reduced density matrix becomes diagonal:

$$\rho_{\alpha_j,\alpha_j'}(R_f, \theta_f, \phi_f) = \frac{\frac{k_{\alpha_j,\beta_j}}{k_0} |f_{\alpha_0,\beta_0 \to \alpha_j,\beta_j}|^2}{d\sigma/d\Omega} \delta_{\alpha_j,\alpha_j'} \\ = \frac{d\sigma_{\alpha_j,\beta_j}/d\Omega}{d\sigma/d\Omega} \delta_{\alpha_j,\alpha_j'}. \quad (16)$$

Therefore, the entanglement entropy $S_{ent}^{int-int}$ can be computed directly from the state-to-state ($d\sigma_{\alpha_j,\beta_j}/d\Omega$) and total ($d\sigma/d\Omega$) differential cross sections

$$S_{ent}^{int-int}(\theta_f, \phi_f) = -\sum_j \left(\frac{d\sigma_{\alpha_j,\beta_j}/d\Omega}{d\sigma/d\Omega}\right) \log_2 \left(\frac{d\sigma_{\alpha_j,\beta_j}/d\Omega}{d\sigma/d\Omega}\right). \quad (17)$$

A concrete example is a collision that only involves spin-exchange $|0\rangle_A |1\rangle_B \leftrightarrow |1\rangle_A |0\rangle_B$ [13]. In this situation, the discrete–discrete entanglement can be expressed in terms of the ratio between the two relevant state-to-state differential cross sections (elastic and spin-exchange transitions) $\gamma = \frac{d\sigma_{01}}{d\Omega} / \frac{d\sigma_{10}}{d\Omega}$:

$$S_{ent}^{int-int}(\theta_f, \phi_f) = \frac{-\gamma}{1+\gamma} \log_2 \left(\frac{\gamma}{1+\gamma}\right) \\ - \frac{1}{1+\gamma} \log_2 \left(\frac{1}{1+\gamma}\right), \quad (18)$$

The entanglement reaches its maximum value when $\gamma = 1$, corresponding to balanced differential cross sections, and decreases as the imbalance between the two state-to-state differential cross sections increases. Note that this expression can also be derived from the qubit–qubit model discussed above (see Supplementary Material S3).

### B. DV-DV entanglement in ultracold Rb + SrF and Rb + Sr$^+$ collisions: Internal state dependence and magnetic tuning

Beyond the two analytical cases considered in the previous section, the evaluation of the entanglement entropy in realistic scattering systems should be performed numerically via diagonalization of the reduced density matrix. Here, we consider the generation of DV-DV entanglement in ultracold Rb + SrF and Rb + Sr$^+$ collisions, both of which are benchmark systems in ultracold molecular dynamics [48, 49] and hybrid ion-atom physics [50–53].

We first quantify the DV-DV entanglement produced in ultracold collisions of SrF molecules with $^{87}$Rb atoms at 100 nK (with details of the S-matrix scattering calculations provided in Supplementary Material S8.) Both Rb atoms and SrF molecules can be prepared by laser cooling and confined in an optical dipole trap [54–57], making Rb + SrF an ideal benchmark system to study the quantum dynamics and control of ultracold atom-molecule collisions [48, 49]. Closely related Rb + CaF [58], Na + NaLi [59], and K + NaK [60] collisions have already been explored experimentally at ultralow temperatures, with rich spectra of magnetic Feshbach resonances observed for Na + NaLi and K + NaK.

In their rovibrational ground state, SrF molecules exhibit two hyperfine manifolds corresponding to the total angular momentum quantum numbers $F = 0, 1$. The ground-state hyperfine structure of $^{87}$Rb atoms consists of two manifolds with $F = 1, 2$. Throughout this work, we label these internal hyperfine states in the order of increasing energy [49], e.g., $|1\rangle_{\text{SrF}} \equiv |F = 0, m = 0\rangle$, $|2/3/4\rangle_{\text{SrF}} \equiv |F = 1, m = -1/0/1\rangle$ for SrF molecule, and $|1/2/3\rangle_{\text{Rb}} \equiv |F = 1, m = 1/0/-1\rangle$, $|4/5/6/7/8\rangle_{\text{Rb}} \equiv |F = 2, m = -2/-1/0/1/2\rangle$ for Rb atom. (Note that states within the same hyperfine manifold are grouped together for compactness; for instance, $|3\rangle_{\text{SrF}}$ denotes the $|F = 1, m = 0\rangle$ state.)

Quantum scattering calculations were performed for several initial internal states, and the complete set of results is provided in Supplementary Material S7. Selected illustrative cases corresponding to the initial states $|2\rangle_{\text{SrF}} |1\rangle_{\text{Rb}}$ and $|3\rangle_{\text{SrF}} |3\rangle_{\text{Rb}}$ are discussed in the following. Note that the collisions are in a double s-wave regime ($\ell = 0 \to \ell' = 0$), so that the differential cross section is independent of the final relative orientation ($\theta_f, \phi_f$) (if taken differently from the initial relative orientation). The amount of post-collision entanglement can therefore be analyzed in terms of the integral cross section The detection distance is taken to be sufficiently large, with $R_f = 10^{10}$ a.u..

For the initial state $|2\rangle_{\text{SrF}} |1\rangle_{\text{Rb}}$, which corresponds to a spin-exchange scenario in which only the final states $|1\rangle_{\text{SrF}} |2\rangle_{\text{Rb}}$ ($\sigma_{21\to 12} = 1843$ Å$^2$) and $|2\rangle_{\text{SrF}} |1\rangle_{\text{Rb}}$ ($\sigma_{21\to 21} = 6892$ Å$^2$) are populated after collision, the numerically obtained entanglement entropy is 0.74, approaching the maximal value of 1. This demonstrates the strong entangling power of spin-exchange collisions for internal states. This numerical result is in excellent agreement with the analytical prediction from Eq. (18) using the cross section ratio $\gamma = 6892/1843 = 3.74$. This case exemplifies a scenario in which the entanglement can be directly inferred from state-to-state cross sections because each internal state of $A$ is correlated with a specific internal state of $B$.

A different scenario is realized for the initial state $|3\rangle_{\text{SrF}} |3\rangle_{\text{Rb}}$, where the states $|1\rangle_{\text{SrF}} |3\rangle_{\text{Rb}}$ ($\sigma_{33\to 13} = 1784$ Å$^2$), $|2\rangle_{\text{SrF}} |2\rangle_{\text{Rb}}$ ($\sigma_{33\to 22} = 746$ Å$^2$) and $|3\rangle_{\text{SrF}} |3\rangle_{\text{Rb}}$ ($\sigma_{33\to 33} = 7671$ Å$^2$) are populated in ultracold Rb + SrF collisions. The calculated DV-DV entanglement entropy is 0.38. In this case, there is no one-to-one correspondence between the internal states of particles $A$ and $B$, and therefore the entanglement cannot be directly inferred from state-to-state cross sections. A complete analysis requires the evaluation of the full reduced density matrix.

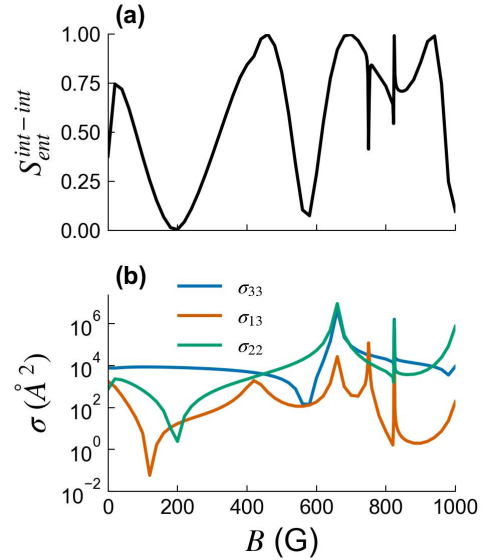

FIG. 2: Magnetic-field control of DV-DV entanglement in ultracold Rb + SrF collisions at 100 nK for the initial state $|3\rangle_{\text{SrF}} |3\rangle_{\text{Rb}}$. (a) DV-DV entanglement entropy between the internal states $S_{ent}^{int-int}$ as a function of the applied magnetic field strength $B$ (b) State-to-state scattering cross sections as a function of the applied magnetic-field strength. The detection distance was fixed at $R = 10^{10}$ a.u.

The quantum dynamics of ultracold atom-atom and atom-molecule collisions is highly tunable through magnetic Feshbach resonances [49, 59–61], leading one to expect that DV–DV entanglement can also be controlled by applying an external magnetic field. To explore this possibility, we computed the entanglement entropy for the initial state $|3\rangle_{\text{SrF}} |3\rangle_{\text{Rb}}$ and magnetic field strengths $B = 0 - 1000$ G (see Fig. 2). The results show substantial tunability of the generated entanglement, which can be continuously varied from zero (corresponding to no entanglement) to unity (corresponding to the formation of a maximally entangled Bell state). Significant variations in entanglement occur in the vicinity of Feshbach resonances, where the inelastic scattering cross sections become comparable to the elastic cross sections. In contrast, the entanglement is negligible in regions where either elastic or inelastic cross section dominates, such as near $B \approx 200$, 600 and 1000 G.

As emphasized above, in the case of ultracold collisions between SrF and Rb, the DV-DV entanglement is independent of the relative orientation of the outgoing particles, and therefore the dependence of the DV-DV entanglement on the relative orientation of the outgoing particles cannot be illustrated with this example. To investigate this dependence, we consider ultracold collisions between a Rb atom and an Sr$^+$ ion at 2 $\mu$K. In this system, the long-range atom–ion interaction, which scales as $1/R^4$, leads to contributions from multiple partial waves (specifically the $s$-, $p$-, and $d$-wave channels). As a result, the amount of entanglement depends on the final relative orientation of the collision products. The hyperfine states considered for $^{87}$Rb are identical to those used above for SrF + Rb collisions. The Sr$^+$ ion is a spin-1/2 system with no hyperfine structure. Accordingly, we denote its spin basis by $|1\rangle_{Sr^+} \equiv |S = 1/2, m = -1/2\rangle$, $|2\rangle_{Sr^+} \equiv |S = 1/2, m = 1/2\rangle$. Details of the S-matrix scattering calculations are provided in Supplementary Material S8.

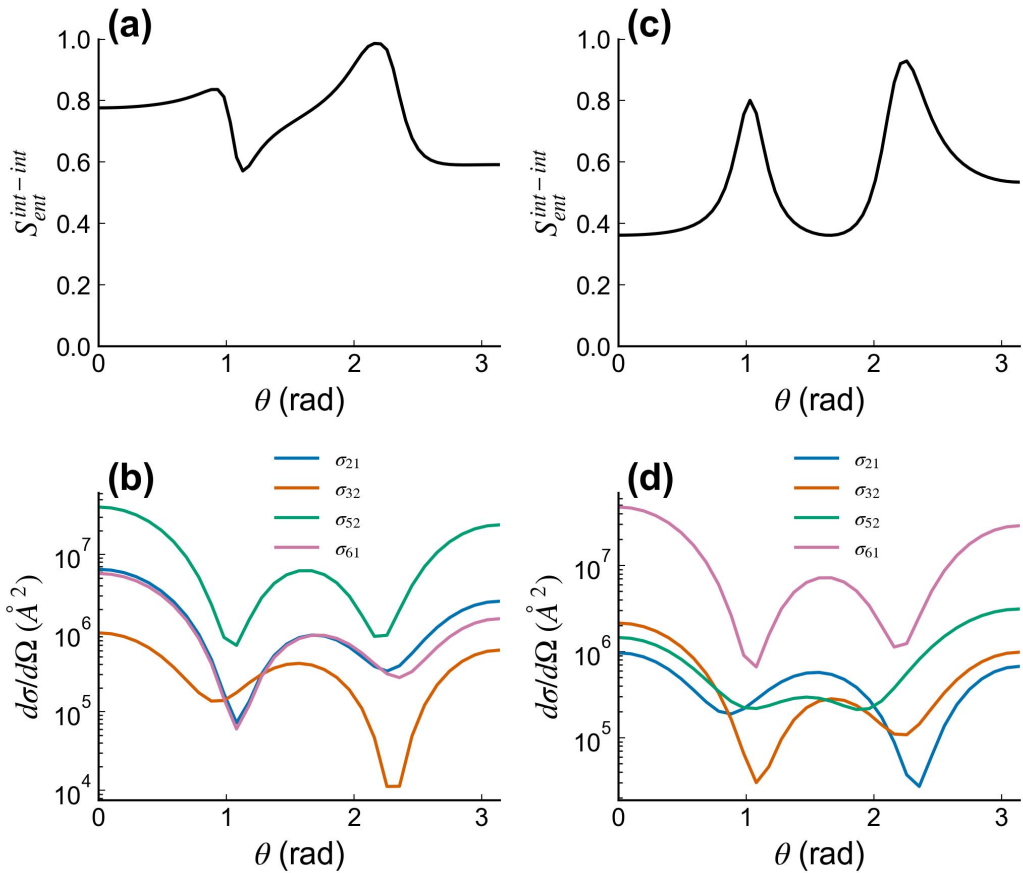

FIG. 3: Angular dependence of the discrete–discrete entanglement generated in Rb-Sr$^+$ collisions for two different initial states: (a) $|5,2\rangle$ and (c) $|6,1\rangle$. The corresponding state-to-state differential cross sections are shown in panels (b) and (d), respectively. The detection distance was fixed at $R = 10^{10}$ a.u.

Figure 3 shows this variation of the entanglement entropy as a function of the scattering angle $\theta_f$ for the initial states $|5\rangle_{\text{Rb}}|2\rangle_{Sr^+}$ and $|6\rangle_{\text{Rb}}|1\rangle_{Sr^+}$ of Rb-Sr$^+$, along with the corresponding state-to-state differential cross sections. For the initial state $|5\rangle_{\text{Rb}}|2\rangle_{Sr^+}$, the entanglement entropy varies from 0.57 at $\theta_f = 1.13$ to 0.99 at $\theta_f = 2.16$. For $|6\rangle_{\text{Rb}}|1\rangle_{Sr^+}$, it ranges from 0.36 at $\theta_f = 1.66$ to 0.93 at $\theta_f = 2.26$. In both cases, two pronounced maxima in the entanglement entropy are observed. These maxima coincide with minima in the elastic differential cross section, which reduce the dominance of elastic scattering and make the elastic and inelastic contributions comparable, thereby enhancing the production of DV-DV entanglement.

For ultracold collisions between SrF and Rb, as well as between Rb and Sr$^+$, no significant dependence of the entanglement on the detection distance $R_f$ was observed. In both cases, only minor oscillations, smaller than 0.1 % of the corresponding values, were detected. This behavior indicates that the $R_f$-dependence often constitutes a negligible effect arising from the postselection procedure.

## IV. CV-CV ENTANGLEMENT

### A. Inelastic collisions

We begin by quantifying the CV–CV entanglement produced in inelastic collisions, which present a simpler case compared to elastic collisions, discussed below. To this end, we fix the final internal states of par-ticles $A$ and $B$ after the collision, treating them as discrete parameters that define a specific scattering channel. Upon postselection on a chosen final internal state $|\alpha_f\rangle_A |\beta_f\rangle_B$, distinct from the initial state $|\alpha_0\rangle_A |\beta_0\rangle_B$, the outgoing scattering wavefunction can be written as (see Supplementary Material S1 and S4):

$$\Psi_{out,\alpha_f,\beta_f}(\vec{r}_A, \vec{r}_B) = \frac{1}{\sqrt{\sigma_{\alpha_f,\beta_f}}} \int d\theta d\phi \sin\theta \sqrt{\frac{k_{\alpha_f,\beta_f}}{k_0}} f_{\alpha_0,\beta_0 \to \alpha_f,\beta_f}(\theta,\phi)$$
$$\times e^{i(\frac{m_A}{M}\vec{K} + k_{\alpha_f,\beta_f}\hat{k}(\theta,\phi)) \cdot \vec{r}_A} e^{i(\frac{m_B}{M}\vec{K} - k_{\alpha_f,\beta_f}\hat{k}(\theta,\phi)) \cdot \vec{r}_B} |\alpha_f\rangle_A |\beta_f\rangle_B, \quad (19)$$

where $\sigma_{\alpha_f,\beta_f}$ is the state-to-state integral cross section for the transition $(\alpha_0,\beta_0) \to (\alpha_f,\beta_f)$. The unit vector $\hat{k}(\theta,\phi)$ specifies the direction of the outgoing relative motion. Note that this expression for the outgoing wavefunction is formulated in the laboratory frame and depends explicitly on the individual particle coordinates $\vec{r}_A$ and $\vec{r}_B$. It therefore differs from the conventional asymptotic form expressed in the center-of-mass frame. The transformation between these two representations is detailed in Supplementary Material S1.

Equation (19) provides the motional wavefunction in the selected internal channel, from which the CV-CV entanglement between the external degrees of freedom of particles $A$ and $B$ can be calculated. It is worth noting that the outgoing scattering state is written as a continuous Schmidt-like decomposition in momentum space. Due to the conservation of total momentum, the direction of the final momentum of particle $A$, $\frac{m_A}{M}\vec{K} + k_{\alpha_f,\beta_f}\hat{k}(\theta,\phi)$, is strictly correlated with that of particle $B$, $\frac{m_B}{M}\vec{K} - k_{\alpha_f,\beta_f}\hat{k}(\theta,\phi)$.

From this expression, we obtain the reduced density matrix associated with the external degrees of freedom of particle $A$ by tracing the full density matrix constructed from the pure state (19) over the external degrees of freedom of particle $B$ (see derivation in Supplementary Material S4):

$$\rho_A(\vec{r}_A, \vec{r}'_A) = \frac{1}{\sigma_{\alpha_f,\beta_f}} \int d\theta d\phi \sin\theta \left(\frac{d\sigma_{\alpha_f,\beta_f}}{d\Omega}\right)$$
$$\times e^{i(\frac{m_A}{M}\vec{K} + k_{\alpha_f,\beta_f}\hat{k}(\theta,\phi)) \cdot \vec{r}_A} e^{-i(\frac{m_A}{M}\vec{K} + k_{\alpha_f,\beta_f}\hat{k}(\theta,\phi)) \cdot \vec{r}'_A} \quad (20)$$

The resultant reduced density matrix is an incoherent mixture of plane waves of fixed magnitude $k_{\alpha_f,\beta_f}$ but varying propagation directions, each weighted by the state-to-state differential cross section $d\sigma_{\alpha_f,\beta_f}/d\Omega$.

The reduced density matrix in Eq. (20) is diagonal in the momentum representation, which enables an analytical evaluation of the CV-CV entanglement entropy (see derivation in Supplementary Material S4):

$$S_{ent,\alpha_f,\beta_f}^{ext-ext} = -\int d\theta d\phi \sin\theta \frac{\frac{d\sigma_{\alpha_f,\beta_f}}{d\Omega}}{\sigma_{\alpha_f,\beta_f}} \log_2\left(\frac{\frac{d\sigma_{\alpha_f,\beta_f}}{d\Omega}}{\sigma_{\alpha_f,\beta_f}}\right). \quad (21)$$

Crucially, this expression establishes a direct connection between the CV-CV entanglement and measurable scattering observables, namely, the state-to-state differential and total cross sections. The entanglement is fully determined by the angular distribution of the state-to-state differential cross section $\frac{d\sigma_{\alpha_f,\beta_f}}{d\Omega}(\theta,\phi)$. A narrowly peaked angular distribution yields weak entanglement, whereas a broad (ideally isotropic) distribution produces high entanglement. For example, if the final state $(\alpha_f,\beta_f)$ is governed exclusively by an $s$-wave contribution in the outgoing collision channel, the angular distribution is isotropic, and the entanglement entropy reaches its maximal value $S_{ent,\alpha_f,\beta_f}^{ext-ext} = \log_2(4\pi) \approx 3.65$. Such conditions arise in the ultracold regime when a double $s$-wave regime occurs for the final state $|\alpha_f\rangle_A |\beta_f\rangle_B$ [13, 62]. Ultracold Rb + SrF collisions provide a clear example of double $s$-wave regime and, therefore, the generated CV-CV entanglements reach its maximal value, $\log_2(4\pi)$, for all the inelastic channels.

To illustrate the influence of the angular scattering distribution on the generation of CV-CV entanglement, we consider the collision of Sr$^+$ ions with Rb atoms. The scaled state-to-state differential cross sections $\sigma_{\alpha_f,\beta_f}^{-1} d\sigma_{\alpha_f,\beta_f}/d\Omega$ for several final channels are presented in Fig. 4(a), while the corresponding CV-CV entanglement values are shown in Fig. 4(b). As discussed above, broader angular distributions result in more strongly entangled final states. Specifically, the largest entanglement is observed for the final states $|2\rangle_{\text{Rb}}|2\rangle_{Sr^+}$ and $|3\rangle_{\text{Rb}}|1\rangle_{Sr^+}$, which exhibit the broadest angular distributions. In contrast, the angular distributions for the final states $|3\rangle_{\text{Rb}}|2\rangle_{Sr^+}$ and $|6\rangle_{\text{Rb}}|1\rangle_{Sr^+}$ are more localized around 0, $\pi/2$ and $\pi$, leading to less final-state entanglement.

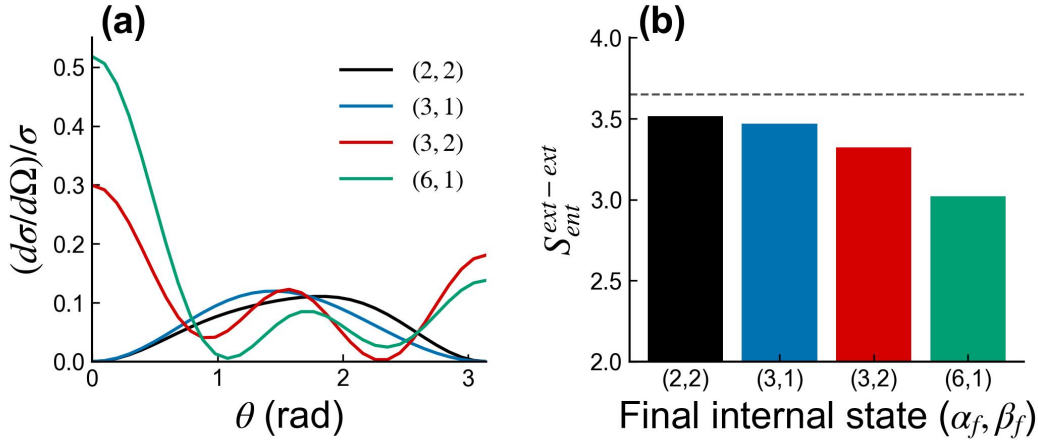

FIG. 4: (a) Scaled state-to-state differential cross sections $(d\sigma_{\alpha_f,\beta_f}/d\Omega)/\sigma_{\alpha_f,\beta_f}$ for Rb + Sr$^+$ collisions from the initial state $|5\rangle_{\text{Rb}} |2\rangle_{Sr^+}$ to four selected final states. (b) CV-CV entanglement generated in these four final scattering channels. The maximum value for the entanglement ($\log_2(4\pi)$) is shown by a horizontal dashed line

### B. Elastic collisions

Until now, the case of elastic collisions ($\alpha_f = \alpha_0$, $\beta_f = \beta_0$) has not been discussed. This situation is more complex because of the presence of the initial (unscattered) component in the total scattering wavefunction. In this case, the outgoing wavefunction contains a term of the form $\delta(\Omega - \Omega_0) + \frac{i}{2\pi} k_0 f_{\alpha_0,\beta_0 \to \alpha_0,\beta_0}(\theta, \phi)$ [46] rather than only the scattered component $\frac{i}{2\pi} k_0 f_{\alpha_0,\beta_0 \to \alpha_0,\beta_0}(\theta, \phi)$. The delta-function contribution, corresponding to the non-interacting part from the incident state, significantly reduces the entanglement, leading to negligible values for an initial plane-wave state. Larger entanglement can be achieved by preparing an initial wavepacket with a broad momentum distribution [28]. This highlights the advantage of postselecting inelastic channels when a goal is to generate large CV-CV entanglement, as opposed to much weaker entanglement produced in elastic collisions.

### C. Cold chemical reactions: F + HD → HF + D

We now explore the CV-CV entanglement generated by the well-known chemical reaction F + HD → HF + D that has played a significant role in the development of chemical reaction dynamics [63], and has been observed to proceed via a tunneling mechanism at temperatures as low as 11 K [64]. The value of the CV-CV entanglement depends on the final rovibrational state $(v', j', m')$ of the HF product, where $m'$ is the projection of $j'$ on a space-fixed quantization axis.

Figure 5(a) shows the product state dependence of the CV-CV entanglement generated in the F + HD $(v=0, j=1, m=0)$ → HF + D chemical reaction at a temperature of 1 K. Only a weak dependence on the final vibrational quantum number is observed, whereas the dependence on the final rotational states $|j', m'\rangle_{\text{HF}}$ is significantly more pronounced. This arises because the final rotational quantum numbers determine the dominant contributing partial waves, through angular-momentum selection rules, and consequently influence the angular distribution of the differential cross section.

Distinct patterns can be identified in Fig. 5(a). The entanglement tends to decrease as $j'$ increases, as shown by the darker shading with increasing $j'$ in Fig. 5(a). Moreover, the dependence on the final magnetic quantum number $m'$ is symmetric: the amount of entanglement associated with the final states $|j', m'\rangle_{\text{HF}}$ and $|j', -m'\rangle_{\text{HF}}$ is identical. For a given value of $j'$, the entanglement entropy attains its lowest values for the fully polarized states $|j', \pm j'\rangle_{\text{HF}}$ and decreases progressively with increasing $j'$. This can be explained by the change of the scaled differential cross section $(d\sigma_{\alpha_f,\beta_f}/d\Omega)/\sigma_{\alpha_f,\beta_f}$ [see Fig. 5(b) for the final states $|0,0\rangle_{\text{HF}}$, $|8,8\rangle_{\text{HF}}$, and $|15,15\rangle_{\text{HF}}$]. As $j'$ increases, the corresponding scaled differential cross section becomes increasingly localized near $\theta = \pi/2$.

Another local minimum in the CV-CV entanglement entropy appears at $m' = 0$ and becomes increasingly pronounced as $j'$ increases [see the darker shading at $m' = 0$ and $j' > 9$ in Fig. 5(a)]. This behavior can again be understood in terms of the scaled differential cross section, as illustrated in Fig. 5(c). As $j'$ increases, the scaled differential cross section corresponding to the final state $|j', 0\rangle_{\text{HF}}$ becomes progressively more localized near $\theta = 0$ and $\theta = \pi$, which leads to a reduction in the entanglement generated. In contrast, the maximum entanglement for a given value of $j'$ occurs at an intermediate value of $m'$. For example, for $j' = 15$ [see Fig. 5(d)], the maximum entanglement is obtained for $m' = 6$, whose scaled differential cross section is comparatively more delocalized. This contrasts with the final states $|15, 15\rangle_{\text{HF}}$ and $|15, 0\rangle_{\text{HF}}$, both of which exhibit

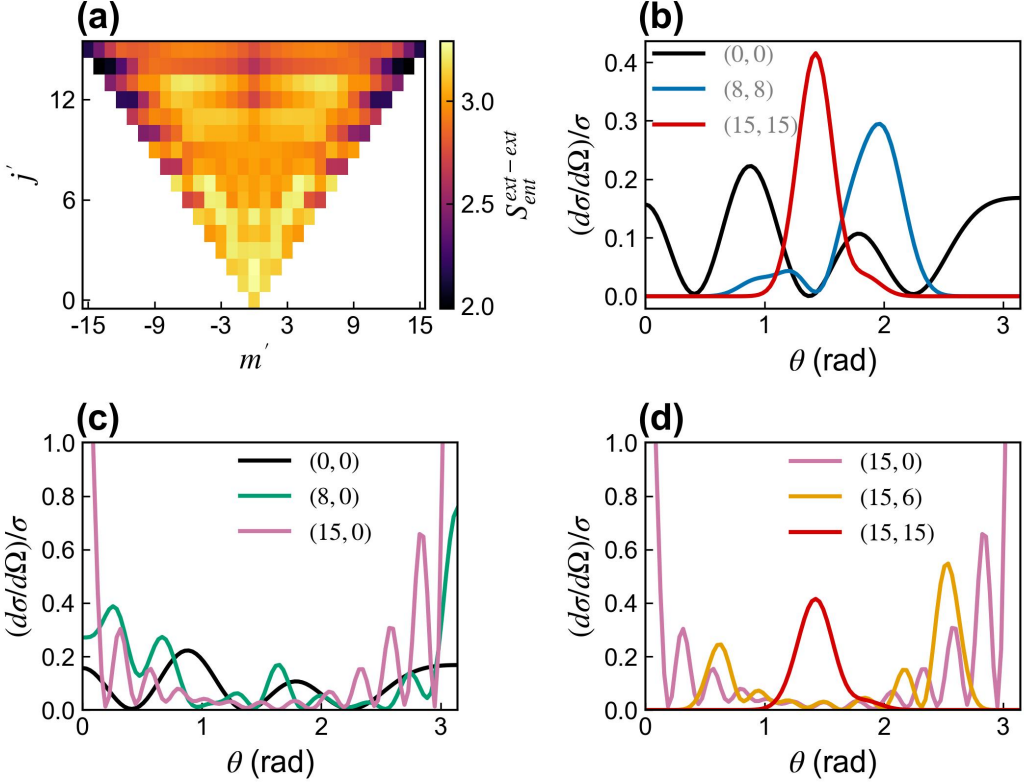

FIG. 5: (a) CV-CV entanglement generated in the chemical reaction F + HD $(v = 0, j = 1, m = 0) \to$ HF$(v', j', m')$ + D, shown as a function of the final product rotational quantum numbers $j'$ and $m'$. The entanglement entropy shown is averaged over all accessible final vibrational states. (b), (c), and (d) Scaled state-to-state differential cross sections $(d\sigma_{\alpha_f,\beta_f}/d\Omega)/\sigma_{\alpha_f,\beta_f}$ for selected final product states $(v', j', m')$. The final vibrational level is fixed as $v' = 3$. The details of the S-matrix scattering calculations are provided in Supplementary Material S8.

localized angular distributions.

## V. HYBRID DV-CV ENTANGLEMENT

### A. Formalism

To quantify the hybrid DV-CV entanglement [44] between the internal and external degrees of freedom of collision products, we explore the outgoing component of the total non-reactive scattering wavefunction [46, 65] (See Supplementary Material S1 for derivation)

$$\Psi_{out}(\vec{r}_A, \vec{r}_B) = \sum_j \int d\theta d\phi \sin\theta \ \tilde{f}_{j_0 \to j}(\theta, \phi)$$
$$\times e^{i(\frac{m_A}{M}\vec{K} + k_j \hat{k}(\theta,\phi)) \cdot \vec{r}_A} e^{i(\frac{m_B}{M}\vec{K} - k_j \hat{k}(\theta,\phi)) \cdot \vec{r}_B} |j\rangle, \quad (22)$$

$$\tilde{f}_{j_0 \to j}(\theta, \phi) = \delta(\Omega - \Omega_0)\delta_{j,j_0} + \frac{i}{2\pi}\sqrt{k_j k_0} f_{j_0 \to j}(\theta, \phi), \quad (23)$$

where $\tilde{f}_{j_0 \to j}(\theta, \phi)$ is the generalized scattering amplitude. It contains the non-scattered and scattered parts. Note that for notational simplicity, we denote the internal states collectively by $|j\rangle$ throughout this section, without distinguishing explicitly the internal states of particles $A$ and $B$. The normalization condition

$$\sum_j \int d\theta d\phi \sin\theta |\tilde{f}_{j_0 \to j}(\theta, \phi)|^2 = 1, \quad (24)$$

ensures conservation of the total outgoing probability i.e., that the sum over all internal channels $j$ and the integral over all scattering directions $(\theta, \phi)$ yields unit total probability for the outgoing component of the scattering wavefunction.

*The asymptotic form of the scattering wavefunction* (22) *highlights a specific type of DV-CV entanglement generated by inelastic and reactive collisions, which we refer to as a "multimode hybrid cat states" by analogy with the regular Schödinger cat states produced in, e.g., trapped-ion experiments [66, 67].*

Indeed, the scattering wavefunction is a superposition of internal states correlated with spherical outgoing waves (equivalently, coherent superpositions of plane-wave momenta of different directions). By contrast, standard hybrid entanglement realized in, e.g., trapped-ion [66] experiments typically involves discrete internal levels (the ion's internal degrees of freedom) entangled with a single or two Gaussian or coherent mode of the ion's center-of-mass motion in the trap. Since each asymptotic momentum direction can be regarded as an independent mode of the relative motion, scattering processes naturally produce hybrid entanglement between the manifold of molecular internal states $|j\rangle$ and a highly multimode continuum subsystem composed of the correlated translational motion of collision products with different momenta. Such multimode hybrid entanglement potentially provides access to larger information-encoding capacity and more potential for quantum technologies.

The degree of hybrid DV–CV entanglement is quantified via the entanglement entropy computed from the eigenvalues of the reduced density matrix, obtained by tracing over the external degrees of freedom of both particles in the total density matrix associated with the outgoing wavefunction (22) (see Supplementary Material S5). The resulting loss of coherence is determined by the degeneracy structure of the internal energy spectrum of the colliding molecules.

If the two internal states $|j\rangle$ and $|j'\rangle$ are non-degenerate, the off-diagonal coherences vanish when taking the trace:

$$\rho_{j,j'} = \int d\theta d\phi \sin\theta |\tilde{f}_{j_0 \to j}(\theta,\phi)|^2 \delta_{j,j'}. \qquad (k_j \neq k_{j'}) \tag{25}$$

If the internal states are degenerate, the coherence between $|j\rangle$ and $|j'\rangle$ is not completely destroyed by the trace over the external degrees of freedom. Instead, it depends on the angular overlap between the corresponding generalized scattering amplitudes $\tilde{f}_j(\theta,\phi)$:

$$\rho_{j,j'} = \int d\theta d\phi \sin\theta \tilde{f}_{j_0 \to j}(\theta,\phi)\tilde{f}^*_{j_0 \to j'}(\theta,\phi). \qquad (k_j = k_{j'}) \tag{26}$$

To obtain finite values for the angular integral, the delta function appearing in the expression for $\tilde{f}_j(\theta,\phi)$ [Eq. (23)] must be regularized. This is achieved by replacing it with a narrow, $L^2$-normalized Gaussian function, which physically corresponds to modeling the initial state as a narrow Gaussian wavepacket rather than an ideal plane wave. The full details of the regularization procedure are provided in the Supplementary Material S6. After regularization, the angular integral simplifies, yielding the following diagonal elements of the reduced density matrix (see Supplementary Material S6 for the derivation):

$$\rho_{jj} = \frac{k_0^2}{4\pi^2}\sigma_{j_0 \to j}, \tag{27}$$

$$\rho_{j_0 j_0} = 1 + \frac{k_0^2}{4\pi^2}(\sigma_{j_0 \to j_0} - \sigma), \tag{28}$$

where $\sigma_{j_0 \to j} = \frac{k_j}{k_0} \int d\theta d\phi \sin\theta |f_{j_0 \to j}(\theta,\phi)|^2$ are the state-to-state integral cross section from the initial state $|j_0\rangle$ to the final state $|j\rangle$ and $\sigma = \sum_j \sigma_{j_0 \to j}$ is the total integral cross section.

The off-diagonal elements of the reduced density matrix take the form (see derivation in Supplementary Material S6):

$$\rho_{j,j'} = \delta_{jj_0}\delta_{j'j_0} + \delta_{jj_0}C^*_{j'} + \delta_{j'j_0}C_j + D_{j,j'}, \tag{29}$$

where

$$C_j = \frac{i}{2\pi}\sqrt{k_0 k_j} f^*_{j_0 \to j}(\Omega_0), \tag{30}$$

$$D_{j,j'} = \frac{k_0}{4\pi^2}\sqrt{k_j k_{j'}} \int d\theta d\phi \sin\theta f_{j_0 \to j}(\theta,\phi) f^*_{j_0 \to j'}(\theta,\phi). \tag{31}$$

The term $C_j$ captures the interference between the non-scattered component of the wavefunction and the forward-scattered amplitude. The term $D_{j,j'}$ corresponds to the angular overlap between the scattering amplitudes.

From the reduced density matrix, the hybrid (internal–external) entanglement entropy is obtained from the eigenvalues $\lambda_i^{int-ext}$ of $\rho_{j,j'}$:

$$S_{ent}^{int-ext} = -\left(\sum_i \lambda_i^{int-ext}\log_2(\lambda_i^{int-ext})\right). \tag{32}$$

If all internal states $|j\rangle$ are non-degenerate (for example, in the presence of a magnetic field), the reduced density matrix becomes diagonal, and the entanglement entropy can be expressed directly in terms of the state-to-state and total integral cross sections:

$$S_{ent}^{int-ext} = -\left(1 + \frac{k_0^2}{4\pi^2}(\sigma_{j_0 \to j_0} - \sigma)\right)\log_2\left(1 + \frac{k_0^2}{4\pi^2}(\sigma_{j_0 \to j_0} - \sigma)\right) - \sum_j \frac{k_0^2}{4\pi^2}\sigma_{j_0 \to j}\log_2\left(\frac{k_0^2}{4\pi^2}\sigma_{j_0 \to j}\right). \tag{33}$$

The first term represents the contribution from the elastic channel, whereas the remaining terms account for the contributions of all inelastic channels. This formulation explicitly shows how hybrid entanglement arises from the distribution of probability among the various internal states through inelastic scattering.

## B. Hybrid CV-DV entanglement in ultracold Rb + SrF and Rb + Sr$^+$ collisions: Internal state dependence and magnetic tuning

Like the DV-DV entanglement, the hybrid DV-CV entanglement can be tuned via an external magnetic field. This control is illustrated for ultracold Rb + SrF collisions in Fig. 6(a) and for ultracold Rb + Sr$^+$ collisions in Fig. 6(b).

Figures 6(c) and 6(e) show the state-to-state cross sections and the diagonal elements of the reduced density matrix given by Eqs. (27) and (28) for ultracold Rb + SrF collisions at 100nK. We observe that the hybrid entanglement entropy is small due to the significant contribution of the forward non-interacting component. This system is an example where the state-to-state integral cross section alone is not an optimal quantity for characterizing the generation of hybrid DV-CV entanglement, since it does not account for the forward, non-interacting contribution. A more informative quantity is given by the diagonal elements of the reduced density matrix, as defined in Eqs. (27) and (28). In this case, hybrid entanglement becomes significant only in the vicinity of scattering resonances, where the elastic and inelastic diagonal elements become comparable in magnitude. In these regions, the hybrid entanglement can vary by several orders of magnitudes, with values tunable from $1.28 \times 10^{-6}$ (at $B = 200$ G) to 0.16 (at $B = 660$ G).

In contrast, for Rb + Sr$^+$ collisions, the forward non-interacting component is less dominant, which enables substantially larger values of hybrid entanglement compared to Rb + SrF collisions. The state-to-state integral cross sections and the diagonal elements of the reduced density matrix [Eqs. (27) and (28)] are shown in Fig. 6(d) and 6(f), respectively. At certain magnetic fields, the hybrid DV-CV entanglement exceeds unity because the diagonal elements $\rho_{jj}$ for four final states have non-negligible values. As in the Rb + SrF case, the hybrid entanglement significantly increases in the vicinity of scattering resonances, where the inelastic contributions become significant. The value of entanglement entropy can be tuned from 0.556 ($B = 580$ G) to 1.96 ($B = 700$ G).

## C. Hybrid DV-CV entanglement production in cold chemical reactions: F + HD → HF + D

Finally, we evaluate the amount of hybrid DV-CV entanglement produced in each possible arrangement of the F + HD chemical reaction: 1) F + HD, 2) HF + D and 3) DF + H. The hybrid entanglement must be defined separately for each reaction arrangement, and a post-selection on the relevant arrangement is required. The formalism described above remains unchanged except for the normalization procedure. Specifically, the reduced density matrices are normalized by an arrangement-dependent constant $N_{arr}$, defined as:

$$N_{arr} = \sum_{j \in arr} \int d\theta d\phi \sin\theta |\tilde{f}_{j_0 \to j}(\theta, \phi)|^2 \quad (34)$$

For reactive arrangements, this normalization constant corresponds to the total reaction cross section, $\sigma_{tot,arr}$. Consequently, Eqs. (27), (28) and (29) for the reduced density matrix take the following arrangement-resolved form:

$$\rho_{jj} = \frac{k_0^2}{4\pi^2} \frac{\sigma_j}{N_{arr}}, \quad (35)$$

$$\rho_{j_0 j_0} = \frac{1 + \frac{k_0^2}{4\pi^2}(\sigma_{j_0} - \sigma)}{N_{arr}}, \quad (36)$$

$$\rho_{j,j'} = \frac{\delta_{jj_0}\delta_{j'j_0} + \delta_{jj_0}C_{j'}^* + \delta_{j'j_0}C_j + D_{j,j'}}{N_{arr}}, \quad (37)$$

Note that terms involving $j_0$ appear only in the non-reactive arrangement (F + HD → F + HD in this case).

We obtain a hybrid entanglement of $1.5 \times 10^{-3}$ for the non-reactive F + HD → F + HD channel, compared with values of 6.1 and 6.6 for the reactive channels F + HD → HF + D and F + HD → DF + H, respectively. The very small value observed in the non-reactive channel is primarily due to the overwhelming dominance of the elastic component (which contains the forward non-interacting and the elastic scattering components). The elastic component is several orders of magnitude larger than that of the inelastic transition $(v = 0, j = 1, m = 0) \to (v' = 0, j' = 0, m' = 0)$. By contrast, the reactive arrangements are not affected by the presence of large elastic components. *This makes reactive processes inherently more favorable than inelastic collisions for generating substantial hybrid entanglement.*

Furthermore, the F + HD chemical reaction populates a broad range of internal states $(v', j', m')$ with different energies, as illustrated in Fig. 7. For the reactive process F + HD → HF + D, 19 final states exhibit non-negligible population (from 0.02 to 0.09). In contrast, for the reactive channel F + HD → DF + H, 11 final states have non-negligible population (between 0.02 and 0.12). The higher entanglement values observed in the second arrangement can be attributed to a broader distribution among these 11 states, with 8 states exhibiting populations between 0.09 and 0.12. By comparison, in the first arrangement, only 4 states have populations between 0.06 and 0.08. These differences are visually evident in Fig. 7: panel (a), corresponding to the first reaction channel, contains a larger number of non-black rectangles, whereas in panel (b), representing the second reactive channel, the non-black rectangles display more uniform coloration, reflecting the broader population distribution.

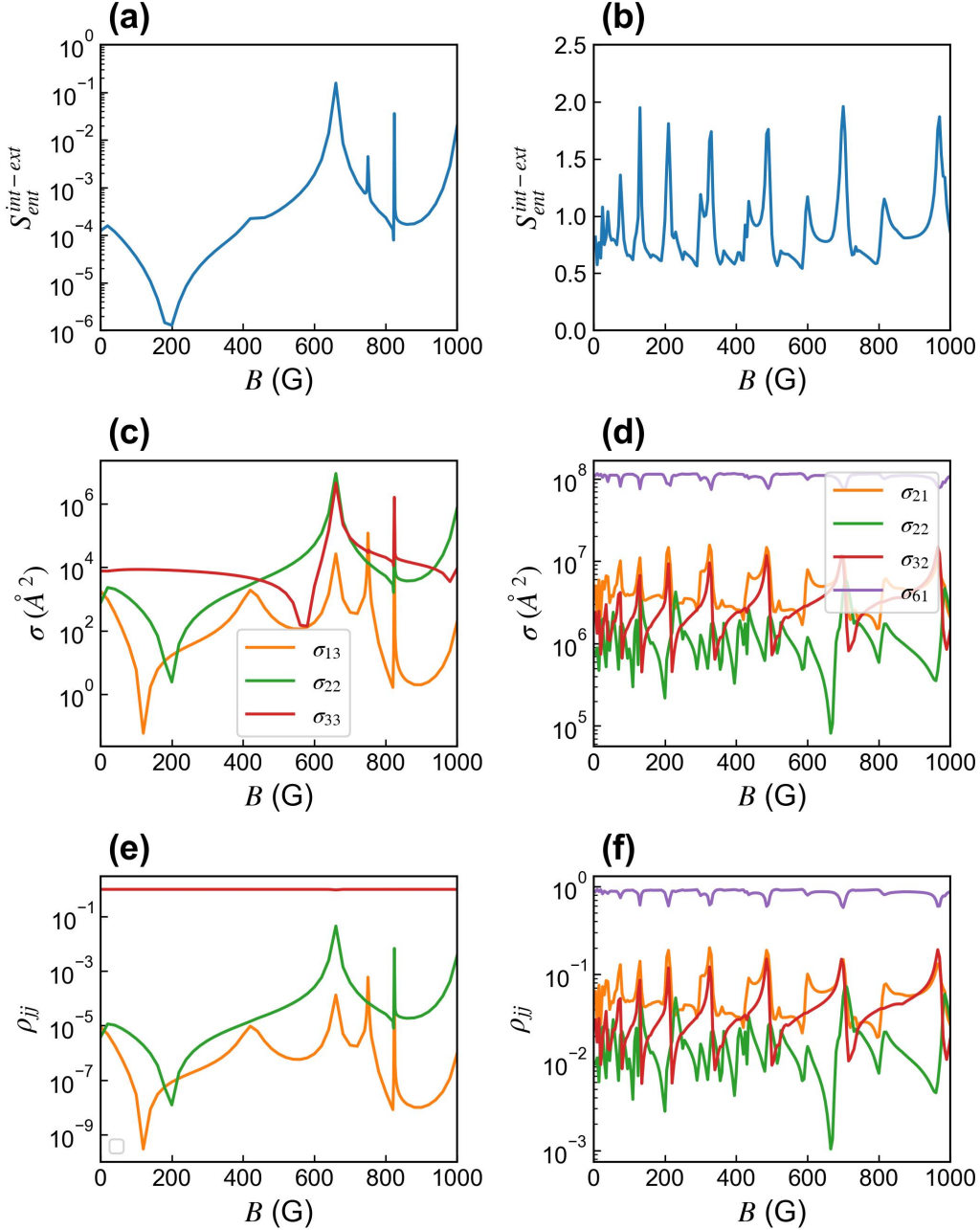

FIG. 6: Hybrid entanglement generated in (a) SrF-Rb with the initial state $|3\rangle_{\text{SrF}} |3\rangle_{\text{Rb}}$, and (b) Rb-Sr$^+$ collisions with the initial state $|6\rangle_{\text{Rb}} |1\rangle_{Sr^+}$. The corresponding state-to-state integral cross sections are presented in panels (c) and (d), respectively, while the diagonal elements of the reduced density matrix are shown in panels (e) and (f), respectively.

## VI. CONCLUSIONS

In this work, we have presented a general framework for quantifying the generation of entanglement in inelastic and reactive molecular collisions, encompassing DV-DV, CV-CV, and hybrid DV-CV entanglement. Each of these types of entanglement arises from correlations between distinct degrees of freedom and may enable different applications in quantum technologies. Using a time-independent scattering formalism, we derived explicit expressions that relate the entanglement entropy to the scattering amplitudes and, consequently, to the S-matrix elements, which can be obtained via coupled-channel calculations. This allows us to quantify the amount of entanglement generated in arbitrary inelastic molecular collisions and chemical reactions. As recently pointed out by Wang and Koch [28], the case of elastic collisions requires extra care, as the amount of entanglement produced depends on the width of the initial wavepacket in momentum space and vanishes for standard plane-wave initial states. For this reason, our framework focuses on the complex dynamics of inelastic and reactive collisions, thereby complementing the results of Ref [28]. Together, these results provide a comprehensive framework for quantifying the generation of entanglement in two-body molecular collisions and chemical reactions. Although the formalism presented in this work is general and is valid at any temperature, we illustrate its versatility for ultracold atom-molecule collisions and chemical reactions (Rb + SrF and F + HD) and atom-ion collisions (Rb + Sr$^+$) using rigorous coupled-channel calculations.

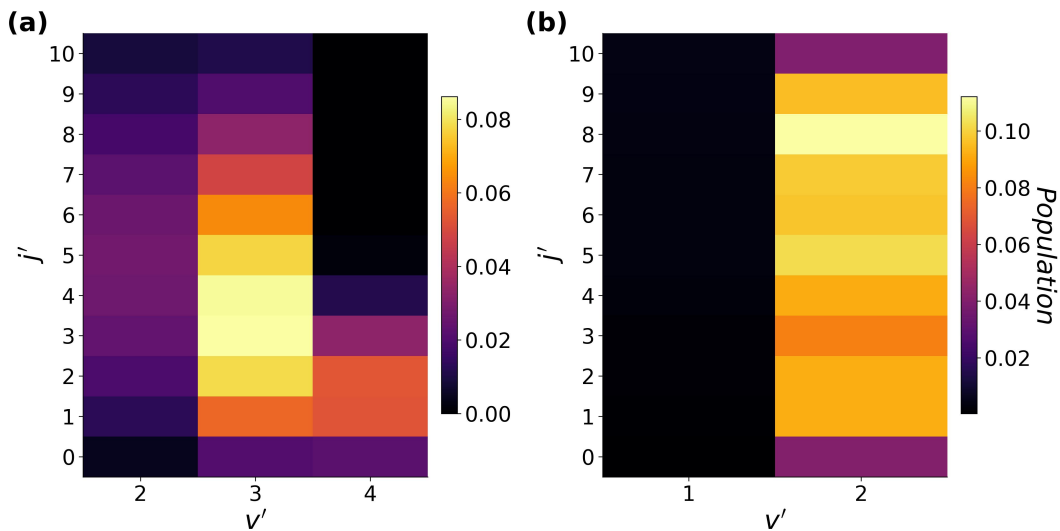

FIG. 7: Population of final rovibrational states (v',j') after the chemical reactions (a) F+HD → HF+D and (b) F+HD → DF+H

Our analysis demonstrates that product-state entanglement generated in ultracold molecular collisions can be actively controlled. In particular, tuning an external magnetic field allows for substantial modulation of DV-DV and hybrid DV-CV entanglement in ultracold collisions near Feshbach resonances, where elastic and inelastic scattering contributions are comparable. This is consistent with recent work on CV-CV entanglement in elastic collisions, which observed enhanced entanglement near shape resonances in the multiple partial wave regime [28]. The quantitative formalism introduced here provides a rigorous metric for optimizing entanglement generation, paving the way for coherent control strategies [68], which will be explored in future work. Overall, this study establishes a unified, computationally accessible framework for predicting and manipulating entanglement in molecular scattering.

While the intriguing question of the experimental detection of product-state entanglement in molecular collisions and chemical reactions is beyond the scope of this work, we note that Wang and Koch have recently considered a number of promising experimental scenarios for observing CV-CV entanglement in elastic collisions, such as velocity map imaging (VMI) with coincidence detection [69, 70], augmented with phase detection via pump-probe spectroscopy [28]. Additionally, full scattering state tomography techniques, such as those developed for detecting photodissociation products [71] and Feshbach resonances in half-collision processes [41] could be extended to two-body collisions, provided that the collision event can be initiated at a well-defined instant of time using, e.g., a sequence of laser pulses [19].

**ACKNOWLEDGEMENTS**

This work was supported by the U.S. Air Force Office of Scientific Research (AFOSR) under Contract No. FA9550-22-1-0361. SciNet computational facilities are gratefully acknowledged.

# Supplementary Material: General framework for quantifying entanglement production in ultracold molecular collisions and chemical reactions

January 16, 2026

## S1: Derivation of the full outgoing wavefunction

The starting point of the entanglement analyses is the full asymptotic outgoing wavefunction expressed in the laboratory frame,

$$\Psi_{\text{out}}(\vec{r}_A, \vec{r}_B) = \sum_{\alpha,\beta} \int d\Omega \, \tilde{f}_{\alpha_0,\beta_0 \to \alpha,\beta}(\Omega) \, e^{i\left(\frac{m_A}{M}\vec{K} + k_{\alpha\beta}\hat{k}(\Omega)\right) \cdot \vec{r}_A} e^{i\left(\frac{m_B}{M}\vec{K} - k_{\alpha\beta}\hat{k}(\Omega)\right) \cdot \vec{r}_B} |\alpha\rangle_A |\beta\rangle_B, \quad (1)$$

where the generalized scattering amplitude is defined as [1]

$$\tilde{f}_{\alpha_0,\beta_0 \to \alpha,\beta}(\Omega) = \delta(\Omega - \Omega_0)\,\delta_{\alpha\alpha_0}\delta_{\beta\beta_0} + \frac{i}{2\pi}\sqrt{k_{\alpha\beta}k_0}\, f_{\alpha_0,\beta_0 \to \alpha,\beta}(\Omega), \quad (2)$$

This expression differs from the conventional asymptotic form of the scattering wavefunction typically presented in the center-of-mass (CM) frame [1, 2],

$$\Psi_{out}(\vec{R}) = [e^{i\vec{k}_0 \vec{R}}]_{outgoing} |\alpha_0, \beta_0\rangle + \sum_{\alpha,\beta} f_{\alpha_0,\beta_0 \to \alpha,\beta}(\hat{R}) \frac{e^{ik_{\alpha,\beta}R}}{R} |\alpha\rangle_A |\beta\rangle_B \quad (3)$$

where $\vec{R}$ denotes the relative coordinate and $\hat{R} = \vec{R}/R$ its direction. In this section, we derive Eq. (13) starting from the CM-frame expression in Eq. (3).

The scattering amplitude can be written in terms of the $S$-matrix elements as

$$f_{\alpha_0,\beta_0 \to \alpha,\beta}(\hat{R}) = \frac{2\pi}{k_0} \sum_{\ell,m_\ell} \sum_{\ell',m'_\ell} i^{\ell - \ell'} Y^*_{\ell,m_\ell}(\hat{k}_0) Y_{\ell',m'_\ell}(\hat{R}) (\delta_{\alpha_0,\beta_0,\ell,m_\ell \to \alpha,\beta,\ell',m'_\ell} - S_{\alpha_0,\beta_0,\ell,m_\ell \to \alpha,\beta,\ell',m'_\ell}) \quad (4)$$

where $\hat{k}_0$ denotes the direction of the incident relative momentum.

The outgoing component of the incident plane wave admits the partial-wave expansion:

$$[e^{ik\hat{k}_0 \vec{r}}]_{outgoing} = -\frac{i2\pi}{k_0} \sum_{\ell,m} Y^*_{\ell,m}(\hat{k}_0) Y_{\ell,m}(\hat{R}) \frac{e^{ik_0 R}}{R} \quad (5)$$

Substituting Eqs. (4) and (5) into Eq. (3) yields the asymptotic outgoing wavefunction in the CM frame in terms of the $S$-matrix,

$$\Psi_{out}(\vec{R}) = \frac{i2\pi}{k_0} \sum_{\alpha,\beta} \sum_{\ell,m} \sum_{\ell',m'} S_{\alpha_0,\beta_0,\ell,m_\ell \to \alpha,\beta,\ell',m'_\ell} Y^*_{\ell,m}(\hat{k}_0) Y_{\ell',m'}(\hat{R}) \frac{e^{ik_{\alpha,\beta}R}}{R} |\alpha\rangle_A |\beta\rangle_B \quad (6)$$

To transform this expression to the laboratory frame, we multiply by the plane wave describing the center-of-mass motion,

$$\Psi_{out}(\vec{r}, \vec{R}_{CM}) = \frac{i2\pi}{k_0} e^{i\vec{K} \cdot \vec{R}_{CM}} \sum_{\alpha,\beta} \sum_{\ell,m} \sum_{\ell',m'} S_{\alpha_0,\beta_0,\ell,m_\ell \to \alpha,\beta,\ell',m'_\ell} Y^*_{\ell,m}(\hat{k}_0) Y_{\ell',m'}(\hat{R}) \frac{e^{ik_{\alpha,\beta}R}}{R} |\alpha\rangle_A |\beta\rangle_B \quad (7)$$

where $\vec{K}$ is the total momentum.

This representation still depends explicitly on both the center-of-mass coordinate $\vec{R}_{CM}$ and the relative coordinate $\vec{R}$. For the purpose of entanglement analyses, it is necessary to rewrite the wavefunction in terms of the individual particle coordinates $\vec{r}_A$ and $\vec{r}B$. For plane waves, this separation is straightforward:

$$e^{i\vec{K}\vec{R}_{CM}} e^{i\vec{k}\vec{R}} = e^{i\vec{k}_A \vec{r}_A} e^{i\vec{k}_B \vec{r}_B} \tag{8}$$

Accordingly, we express the outgoing spherical wave in the relative coordinate as a continuous superposition of plane waves,

$$\frac{e^{ik_{\alpha,\beta}R}}{R} Y_{\ell,m}(\hat{R}) = \int d\hat{k}\, Y_{\ell,m}(\hat{k}) e^{ik_{\alpha,\beta}\hat{k}\vec{R}} \tag{9}$$

which corresponds to the inverse partial-wave expansion.

Substituting Eq. (9) into Eq. (7) leads to Eq. (7):

$$\Psi_{out}(\vec{r}, \vec{R}_{CM}) = e^{i\vec{K}\vec{R}_{CM}} \int d\hat{k}\, \left( \frac{i2\pi}{k_0} \sum_{\alpha,\beta} \sum_{\ell,m} \sum_{\ell',m'} S_{\alpha_0,\beta_0,\ell,m_\ell \to \alpha,\beta,\ell',m'_\ell} Y^*_{\ell,m}(\hat{k}_0) Y_{\ell',m'}(\hat{k}) \right) e^{ik_{\alpha,\beta}\hat{k}\cdot\vec{R}} |\alpha\rangle_A |\beta\rangle_B \tag{10}$$

The product of plane waves now allows the introduction of the individual particle coordinates. Using the relations $\vec{k}_A = \frac{m_A}{M}\vec{K} + k_{\alpha\beta}\hat{k}$ and $\vec{k}_B = \frac{m_B}{M}\vec{K} - k_{\alpha\beta}\hat{k}$, the outgoing wavefunction can finally be written as:

$$\Psi_{out}(\vec{r}_A, \vec{r}_B) = \sum_{\alpha,\beta} \int d\Omega\, \tilde{f}_{\alpha_0,\beta_0 \to \alpha,\beta}(\Omega)\, e^{i\left(\frac{m_A}{M}\vec{K} + k_{\alpha\beta}\hat{k}(\Omega)\right)\cdot\vec{r}_A} e^{i\left(\frac{m_B}{M}\vec{K} - k_{\alpha\beta}\hat{k}(\Omega)\right)\cdot\vec{r}_B} |\alpha\rangle_A |\beta\rangle_B, \tag{11}$$

where the generalized scattering amplitude $\tilde{f}_{\alpha_0,\beta_0 \to \alpha,\beta}(\Omega)$ is expressed in terms of the $S$-matrix elements as:

$$\tilde{f}_{\alpha_0,\beta_0 \to \alpha,\beta}(\Omega) = \frac{i2\pi}{k_0} \sum_{\alpha,\beta} \sum_{\ell,m} \sum_{\ell',m'} S_{\alpha_0,\beta_0,\ell,m_\ell \to \alpha,\beta,\ell',m'_\ell} Y^*_{\ell,m}(\hat{k}_0) Y_{\ell',m'}(\hat{k}) \tag{12}$$

## S2: Derivation of the expression for the DV-DV entanglement entropy

The full outgoing wavefunction after the collision can be written as (see S1 for the derivation):

$$\Psi_{out}(\vec{r}_A, \vec{r}_B) = \sum_{\alpha,\beta} \int d\Omega\, \tilde{f}_{\alpha_0,\beta_0 \to \alpha,\beta}(\Omega)\, e^{i\left(\frac{m_A}{M}\vec{K} + k_{\alpha\beta}\hat{k}(\Omega)\right)\cdot\vec{r}_A} e^{i\left(\frac{m_B}{M}\vec{K} - k_{\alpha\beta}\hat{k}(\Omega)\right)\cdot\vec{r}_B} |\alpha\rangle_A |\beta\rangle_B, \tag{13}$$

where $M = m_A + m_B$ is the total mass, $\vec{K}$ is the center-of-mass momentum, and $\hat{k}(\Omega)$ denotes the unit vector in the scattering direction $\Omega = (\theta, \phi)$. The generalized scattering amplitude $\tilde{f}$ contains both the incoming and scattered contributions [1],

$$\tilde{f}_{\alpha_0,\beta_0 \to \alpha,\beta}(\Omega) = \delta(\Omega - \Omega_0)\, \delta_{\alpha\alpha_0} \delta_{\beta\beta_0} + \frac{i}{2\pi} \sqrt{k_{\alpha\beta} k_0}\, f_{\alpha_0,\beta_0 \to \alpha,\beta}(\Omega), \tag{14}$$

with $f_{\alpha_0,\beta_0 \to \alpha,\beta}$ the scattering amplitude and $k_0$ the incoming relative wave number. A coincidence detection scheme is considered in which the two particles are detected at a large interparticle separation $R_f$ in the opposite directions. The postselected detection positions are

$$\vec{r}_A = \frac{R_f}{2}\, \hat{u}(\theta_f, \phi_f), \qquad \vec{r}_B = -\frac{R_f}{2}\, \hat{u}(\theta_f, \phi_f), \tag{15}$$

where $\hat{u}(\theta_f, \phi_f)$ denotes the detection direction. Under this postselection, the outgoing wavefunction becomes

$$\Psi_{out}(R_f, \theta_f, \phi_f) = \sum_{\alpha,\beta} \int d\Omega'\, \tilde{f}_{\alpha_0,\beta_0 \to \alpha,\beta}(\Omega')$$

$$\times \exp\left[i\left(\frac{m_A}{M}\vec{K} + k_{\alpha\beta}\hat{k}(\Omega')\right) \cdot \frac{R_f}{2}\hat{u}\right] \exp\left[i\left(\frac{m_B}{M}\vec{K} - k_{\alpha\beta}\hat{k}(\Omega')\right) \cdot \left(-\frac{R_f}{2}\hat{u}\right)\right] |\alpha\rangle_A |\beta\rangle_B. \tag{16}$$

Combining the exponential factors yields

$$\Psi_{\text{out}}(R_f, \theta_f, \phi_f) = e^{i\frac{R_f}{2}\left(\frac{m_A - m_B}{M}\right)\vec{K}\cdot\hat{u}} \sum_{\alpha,\beta} \int d\Omega'\, \tilde{f}_{\alpha_0,\beta_0 \to \alpha,\beta}(\Omega')\, e^{ik_{\alpha\beta} R_f\, \hat{k}(\Omega')\cdot\hat{u}} \ket{\alpha}_A \ket{\beta}_B. \tag{17}$$

In the asymptotic limit $R_f \to \infty$, the angular integral in Eq. (17) becomes highly oscillatory. Defining the phase function

$$\Phi(\Omega') = k_{\alpha\beta} R_f\, \hat{k}(\Omega') \cdot \hat{u}(\theta_f, \phi_f), \tag{18}$$

the stationary-phase condition $\nabla_{\Omega'}\Phi = 0$ is satisfied when

$$\hat{k}(\Omega') = \hat{u}(\theta_f, \phi_f). \tag{19}$$

Expanding the phase to second order around the stationary point and applying the two-dimensional stationary-phase approximation on the unit sphere yields, up to an overall channel-independent prefactor,

$$\int d\Omega'\, \tilde{f}_{\alpha_0,\beta_0 \to \alpha,\beta}(\Omega')\, e^{ik_{\alpha\beta} R_f\, \hat{k}(\Omega')\cdot\hat{u}} \simeq \frac{\sqrt{k_{\alpha\beta} k_0}}{R_f}\, f_{\alpha_0,\beta_0 \to \alpha,\beta}(\theta_f, \phi_f)\, e^{ik_{\alpha\beta} R_f}. \tag{20}$$

Only the scattering contribution to $\tilde{f}_{\alpha_0,\beta_0 \to \alpha,\beta}$ is retained because the coincidence detection angles $(\theta_f, \phi_f)$ are chosen to be different from the incident direction $\Omega_0$

The postselected outgoing wavefunction in the asymptotic regime therefore takes the form

$$\Psi_{\text{out}}(R_f, \theta_f, \phi_f) = e^{i\frac{R_f}{2}\left(\frac{m_A - m_B}{M}\right)\vec{K}\cdot\hat{u}} \sum_{\alpha,\beta} \frac{\sqrt{k_{\alpha\beta} k_0}}{R_f}\, f_{\alpha_0,\beta_0 \to \alpha,\beta}(\theta_f, \phi_f)\, e^{ik_{\alpha\beta} R_f} \ket{\alpha}_A \ket{\beta}_B. \tag{21}$$

Conditioned on coincidence detection at $(R_f, \theta_f, \phi_f)$, the normalization factor is

$$N^2 = \sum_{\alpha,\beta} \frac{k_{\alpha\beta} k_0}{R_f^2}\, |f_{\alpha_0,\beta_0 \to \alpha,\beta}(\theta_f, \phi_f)|^2 = \frac{k_0^2}{R_f^2}\, \frac{d\sigma(\theta_f, \phi_f)}{d\Omega}, \tag{22}$$

where $d\sigma/d\Omega$ is the total differential cross section. The normalized conditional internal state thus reads

$$\Psi_{\text{out}}(R_f, \theta_f, \phi_f) = \frac{e^{i\frac{R_f}{2}\left(\frac{m_A - m_B}{M}\right)\vec{K}\cdot\hat{u}}}{\sqrt{\frac{d\sigma(\theta_f, \phi_f)}{d\Omega}}} \sum_{\alpha,\beta} \sqrt{\frac{k_{\alpha\beta}}{k_0}}\, f_{\alpha_0,\beta_0 \to \alpha,\beta}(\theta_f, \phi_f)\, e^{ik_{\alpha\beta} R_f} \ket{\alpha}_A \ket{\beta}_B. \tag{23}$$

The corresponding density matrix is

$$\rho_{\alpha\beta,\alpha'\beta'}(R_f) = \frac{\sqrt{k_{\alpha\beta} k_{\alpha'\beta'}}\, f_{\alpha_0,\beta_0 \to \alpha,\beta} f^*_{\alpha_0,\beta_0 \to \alpha',\beta'}\, e^{i(k_{\alpha\beta} - k_{\alpha'\beta'})R_f}}{k_0 \frac{d\sigma}{d\Omega}}. \tag{24}$$

Tracing over the internal states of particle $B$ yields the reduced density matrix

$$\rho^{\text{out}}_{\alpha\alpha'}(R_f, \theta_f, \phi_f) = \frac{\sum_\beta \sqrt{k_{\alpha\beta} k_{\alpha'\beta}}\, f_{\alpha_0,\beta_0 \to \alpha,\beta} f^*_{\alpha_0,\beta_0 \to \alpha',\beta}\, e^{i(k_{\alpha\beta} - k_{\alpha'\beta})R_f}}{k_0 \frac{d\sigma(\theta_f, \phi_f)}{d\Omega}}. \tag{25}$$

Diagonalization of $\rho^{\text{out}}_{\alpha\alpha'}$ yields the eigenvalues $\{\lambda_i^{\text{int}-\text{int}}\}$, from which the internal–internal entanglement entropy is obtained as

$$S^{\text{int}-\text{int}}_{\text{ent}}(\theta_f, \phi_f) = -\sum_i \lambda_i^{\text{int}-\text{int}} \log_2\!\left(\lambda_i^{\text{int}-\text{int}}\right). \tag{26}$$

## S3: Entanglement in spin-exchange collisions: The two-qubit collision model

The entanglement entropy associated with DV-DV (internal–internal) degrees of freedom for a two-qubit collision is defined as

$$S_{ent}^{int-int}(R_f, \theta_f, \phi_f) = -\lambda_1^{int-int} \log_2(\lambda_1^{int-int}) - \lambda_2^{int-int} \log_2(\lambda_2^{int-int}) \tag{27}$$

where $\lambda_{1,2}^{int-int}$ are the eigenvalues of the reduced density matrix. These eigenvalues are given by:

$$\lambda_{1,2}^{int-int} = \frac{1 \pm \sqrt{1 - 4\Delta}}{2}, \tag{28}$$

with:

$$\Delta = \frac{1}{\left(\frac{d\sigma}{d\Omega}\right)^2} \left(\mathcal{D}_1 - \mathcal{D}_2 \cos(\Phi)\right). \tag{29}$$

$$\mathcal{D}_1 = \frac{d\sigma_{11}}{d\Omega}\frac{d\sigma_{00}}{d\Omega} + \frac{d\sigma_{01}}{d\Omega}\frac{d\sigma_{10}}{d\Omega} \tag{30}$$

$$\mathcal{D}_2 = 2\sqrt{\frac{d\sigma_{00}}{d\Omega}\frac{d\sigma_{01}}{d\Omega}\frac{d\sigma_{10}}{d\Omega}\frac{d\sigma_{11}}{d\Omega}} \tag{31}$$

$$\Phi = \chi_{00} + \chi_{11} - \chi_{01} - \chi_{10} + (k_{00} + k_{11} - k_{01} - k_{10})R_f \tag{32}$$

In the regime where only exchange transitions, $|01\rangle \leftrightarrow |10\rangle$ and elastic processes contribute and the initial state is either $|01\rangle$ or $|10\rangle$, the only non-vanishing state-to-state differential cross sections are $\frac{d\sigma_{01}}{d\Omega}$ and $\frac{d\sigma_{10}}{d\Omega}$. Under these conditions, the quantity $\Delta$ simplifies to

$$\Delta = \frac{\frac{d\sigma_{01}}{d\Omega}\frac{d\sigma_{10}}{d\Omega}}{\left(\frac{d\sigma_{01}}{d\Omega} + \frac{d\sigma_{10}}{d\Omega}\right)^2}. \tag{33}$$

Introducing the ratio $\gamma = \frac{\frac{d\sigma_{01}}{d\Omega}}{\frac{d\sigma_{10}}{d\Omega}}$, the above expression can be written as:

$$\Delta = \frac{\gamma}{(1+\gamma)^2}. \tag{34}$$

Using the identity $1 - \frac{4\gamma}{1+\gamma^2} = \frac{(1-\gamma)^2}{(1+\gamma)^2}$, the eigenvalues reduce to:

$$\lambda_1 = \frac{1}{1+\gamma} \tag{35}$$

$$\lambda_2 = \frac{\gamma}{1+\gamma} \tag{36}$$

Substitution of these eigenvalues into the definition of the entanglement entropy yields the expression for the entanglement generated by an exchange process,

$$S_{ent}^{int-int}(\theta_f, \phi_f) = \frac{-\gamma}{1+\gamma} \log_2\left(\frac{\gamma}{1+\gamma}\right) - \frac{1}{1+\gamma} \log_2\left(\frac{1}{1+\gamma}\right), \tag{37}$$

## S4: Derivation of the expression for the CV-CV entanglement entropy

We start with the full outgoing wavefunction (see S1 for the derivation) :

$$\Psi_{out}(\vec{r}_A, \vec{r}_B) = \sum_j \int d\theta d\phi \sin\theta \; \tilde{f}_{j_0 \to j}(\theta, \phi) e^{i(\frac{m_A}{M}\vec{K} + k_j\hat{k}(\theta,\phi))\vec{r}_A} e^{i(\frac{m_B}{M}\vec{K} - k_j\hat{k}(\theta,\phi))\vec{r}_B} |\alpha\rangle_A |\beta\rangle_B, \tag{38}$$

Upon postselection on a chosen final internal state $|\alpha_f, \beta_f\rangle$, distinct from the initial state $|\alpha_0, \beta_0\rangle$, the outgoing scattering wavefunction can be written as:

$$\Psi_{out,\alpha_f,\beta_f}(\vec{r}_A, \vec{r}_B) = \int d\theta d\phi \sin\theta \sqrt{k_{\alpha_f,\beta_f} k_0} f_{\alpha_0,\beta_0 \to \alpha_f,\beta_f}(\theta, \phi) \\ \times e^{i(\frac{m_A}{M}\vec{K}+k_{\alpha_f,\beta_f}\hat{k}(\theta,\phi))\vec{r}_A} e^{i(\frac{m_B}{M}\vec{K}-k_{\alpha_f,\beta_f}\hat{k}(\theta,\phi))\vec{r}_B} |\alpha_f\rangle_A |\beta_f\rangle_B,\tag{39}$$

Conditioned on the detection of the internal states $|\alpha_f, \beta_f\rangle$, the normalization factor is:

$$N^2 = k_{\alpha_f,\beta_f} k_0 \int d\theta d\phi \sin(\theta) |f_{\alpha_0,\beta_0 \to \alpha_f,\beta_f}(\theta,\phi)|^2 = k_0^2 \sigma_{\alpha_f,\beta_f} \tag{40}$$

where $\sigma_{\alpha_f,\beta_f}$ is the state-to-state integral cross section. The normalized postselected wavefunction thus read:

$$\Psi_{out,\alpha_f,\beta_f}(\vec{r}_A, \vec{r}_B) = \frac{1}{\sqrt{\sigma_{\alpha_f,\beta_f}}} \int d\theta d\phi \sin\theta \sqrt{\frac{k_{\alpha_f,\beta_f}}{k_0}} f_{\alpha_0,\beta_0 \to \alpha_f,\beta_f}(\theta, \phi) \\ \times e^{i(\frac{m_A}{M}\vec{K}+k_{\alpha_f,\beta_f}\hat{k}(\theta,\phi))\vec{r}_A} e^{i(\frac{m_B}{M}\vec{K}-k_{\alpha_f,\beta_f}\hat{k}(\theta,\phi))\vec{r}_B} |\alpha_f\rangle_A |\beta_f\rangle_B,\tag{41}$$

From this wavefunction, the two-particle density matrix is obtained as

$$\rho_{out}(\vec{r}_A, \vec{r}_B; \vec{r'}_A, \vec{r'}_B) = \Psi_{out,\alpha_f,\beta_f}(\vec{r}_A, \vec{r}_B) \Psi^*_{out,\alpha_f,\beta_f}(\vec{r'}_A, \vec{r'}_B) \tag{42}$$

$$\rho_{out}(\vec{r}_A, \vec{r}_B; \vec{r'}_A, \vec{r'}_B) = \frac{1}{\sigma_{\alpha_f,\beta_f}} \int d\theta d\phi d\theta' d\phi' \sin\theta \sin\theta' \frac{k_{\alpha_f,\beta_f}}{k_0} f_{\alpha_0,\beta_0 \to \alpha_f,\beta_f}(\theta,\phi) f^*_{\alpha_0,\beta_0 \to \alpha_f,\beta_f}(\theta',\phi') \\ e^{i(\frac{m_A}{M}\vec{K}+k_{\alpha_f,\beta_f}\hat{k}(\theta,\phi))\vec{r}_A} e^{i(\frac{m_B}{M}\vec{K}-k_{\alpha_f,\beta_f}\hat{k}(\theta,\phi))\vec{r}_B} \\ e^{-i(\frac{m_A}{M}\vec{K}+k_{\alpha_f,\beta_f}\hat{k}'(\theta',\phi'))\vec{r'}_A} e^{-i(\frac{m_B}{M}\vec{K}-k_{\alpha_f,\beta_f}\hat{k}'(\theta',\phi'))\vec{r'}_B} \tag{43}$$

To quantify the entanglement generated by the collision, one constructs the reduced density matrix of particle $A$ by tracing out the external degrees of freedom of particle $B$:

$$\rho_A(\vec{r}_A, \vec{r'}_A) = \int d\vec{r}_B \rho_{out}(\vec{r}_A, \vec{r}_B; \vec{r'}_A, \vec{r}_B) \tag{44}$$

$$\rho_A(\vec{r}_A, \vec{r'}_A) = \frac{1}{\sigma_{\alpha_f,\beta_f}} \int d\theta d\phi d\theta' d\phi' \sin\theta \sin\theta' \frac{k_{\alpha_f,\beta_f}}{k_0} f_{\alpha_0,\beta_0 \to \alpha_f,\beta_f}(\theta,\phi) f^*_{\alpha_0,\beta_0 \to \alpha_f,\beta_f}(\theta',\phi') \\ \times e^{i(\frac{m_A}{M}\vec{K}+k_{\alpha_f,\beta_f}\hat{k}(\theta,\phi))\vec{r}_A} e^{-i(\frac{m_A}{M}\vec{K}+k_{\alpha_f,\beta_f}\hat{k}'(\theta',\phi'))\vec{r'}_A} \\ \times \int d\vec{r}_B e^{i(k_{\alpha_f,\beta_f}\hat{k}'(\theta',\phi')-k_{\alpha_f,\beta_f}\hat{k}(\theta,\phi))\vec{r}_B} \tag{45}$$

The integral over $\vec{r}_B$ enforces equality of scattering directions,

$$\int d\vec{r}_B, e^{i[k_{\alpha_f,\beta_f}\hat{k}'(\theta',\phi')-k_{\alpha_f,\beta_f}\hat{k}(\theta,\phi)]\cdot\vec{r}_B} = \delta(\theta-\theta'), \delta(\phi-\phi'), \tag{46}$$

so that only coincident angular components survive. Off-diagonal coherences in the angular momentum space vanish, and the reduced density matrix of $A$ reduces to an incoherent mixture of plane-waves states:

$$\rho_A(\vec{r}_A, \vec{r'}_A) = \frac{1}{\sigma_{\alpha_f,\beta_f}} \int d\theta d\phi \sin\theta \frac{k_{\alpha_f,\beta_f}}{k_0} |f_{\alpha_0,\beta_0 \to \alpha_f,\beta_f}(\theta,\phi)|^2 e^{i(\frac{m_A}{M}\vec{K}+k_{\alpha_f,\beta_f}\hat{k}(\theta,\phi))\vec{r}_A} e^{-i(\frac{m_A}{M}\vec{K}+k_{\alpha_f,\beta_f}\hat{k}(\theta,\phi))\vec{r'}_A} \tag{47}$$

The reduced density matrix is transformed into the momentum representation according to:

$$\rho_A(\vec{p}_A, \vec{p'}_A) = \int d\vec{r}_A d\vec{r'}_A e^{-i\vec{p}_A\cdot\vec{r}_A} \rho_A(\vec{r}_A, \vec{r'}_A) e^{i\vec{p'}_A\cdot\vec{r'}_A}. \tag{48}$$

Substituting Eq. (47) into this expression yields:

$$\rho_A(\vec{p}_A, \vec{p}_A') = \frac{1}{\sigma_{\alpha_f,\beta_f}} \int d\theta d\phi \sin\theta \, \frac{k_{\alpha_f,\beta_f}}{k_0} \left|f_{\alpha_0,\beta_0 \to \alpha_f,\beta_f}(\theta,\phi)\right|^2$$
$$\times \int d\vec{r}_A \, e^{i(\frac{m_A}{M}\vec{K}+k_{\alpha_f,\beta_f}\hat{k}(\theta,\phi)-\vec{p}_A)\cdot\vec{r}_A} \int d\vec{r}_A' \, e^{-i(\frac{m_A}{M}\vec{K}+k_{\alpha_f,\beta_f}\hat{k}(\theta,\phi)-\vec{p}_A')\cdot\vec{r}_A'}. \quad (49)$$

The standard Fourier identity

$$\int d\vec{r} \, e^{i\vec{q}\cdot\vec{r}} = (2\pi)^3 \delta^{(3)}(\vec{q}), \quad (50)$$

yields

$$\int d\vec{r}_A \, e^{i(\frac{m_A}{M}\vec{K}+k_{\alpha_f,\beta_f}\hat{k}(\theta,\phi)-\vec{p}_A)\cdot\vec{r}_A} = (2\pi)^3 \delta^{(3)}\left(\vec{p}_A - (\frac{m_A}{M}\vec{K}+k_{\alpha_f,\beta_f}\hat{k}(\theta,\phi))\right), \quad (51)$$

$$\int d\vec{r}_A' \, e^{-i(\frac{m_A}{M}\vec{K}+k_{\alpha_f,\beta_f}\hat{k}(\theta,\phi)-\vec{p}_A')\cdot\vec{r}_A'} = (2\pi)^3 \delta^{(3)}\left(\vec{p}_A' - (\frac{m_A}{M}\vec{K}+k_{\alpha_f,\beta_f}\hat{k}(\theta,\phi))\right). \quad (52)$$

Substituting these results back into the expression for $\rho_A(\vec{p}_A, \vec{p}_A')$ leads to

$$\rho_A(\vec{p}_A, \vec{p}_A') = \frac{(2\pi)^6}{\sigma_{\alpha_f,\beta_f}} \int d\theta d\phi \sin\theta \, \frac{k_{\alpha_f,\beta_f}}{k_0} \left|f_{\alpha_0,\beta_0 \to \alpha_f,\beta_f}(\theta,\phi)\right|^2$$
$$\times \delta^{(3)}\left(\vec{p}_A - (\frac{m_A}{M}\vec{K}+k_{\alpha_f,\beta_f}\hat{k}(\theta,\phi))\right) \delta^{(3)}\left(\vec{p}_A' - (\frac{m_A}{M}\vec{K}+k_{\alpha_f,\beta_f}\hat{k}(\theta,\phi))\right). \quad (53)$$

The product of the two Dirac delta functions can be recast to emphasize the diagonality of the reduced density matrix in momentum space:

$$\rho_A(\vec{p}_A, \vec{p}_A') = \delta^{(3)}(\vec{p}_A - \vec{p}_A') \frac{(2\pi)^6}{\sigma_{\alpha_f,\beta_f}} \int d\theta d\phi \sin\theta \, \frac{k_{\alpha_f,\beta_f}}{k_0} \left|f_{\alpha_0,\beta_0 \to \alpha_f,\beta_f}(\theta,\phi)\right|^2 \delta^{(3)}\left(\vec{p}_A - (\frac{m_A}{M}\vec{K}+k_{\alpha_f,\beta_f}\hat{k}(\theta,\phi))\right) \quad (54)$$

Thus, the reduced density matrix is diagonal in the momentum representation. Consequently, the entanglement entropy can be directly expressed in terms of the eigenvalues as

$$S_{ent,\alpha_f,\beta_f}^{ext-ext} = -\int d\theta d\phi \sin\theta \, \frac{\frac{d\sigma_{\alpha_f,\beta_f}}{d\theta d\phi}}{\sigma_{\alpha_f,\beta_f}} \log_2 \left(\frac{\frac{d\sigma_{\alpha_f,\beta_f}}{d\theta d\phi}}{\sigma_{\alpha_f,\beta_f}}\right). \quad (55)$$

## S5: Derivation of the expression for the DV-CV entanglement entropy

We start with the full outgoing wavefunction (see S1 for the derivation):

$$\Psi_{out}(\vec{r}_A, \vec{r}_B) = \sum_j \int d\theta d\phi \sin\theta \, \tilde{f}_{j_0 \to j}(\theta,\phi) e^{i(\frac{m_A}{M}\vec{K}+k_j\hat{k}(\theta,\phi))\vec{r}_A} e^{i(\frac{m_B}{M}\vec{K}-k_j\hat{k}(\theta,\phi))\vec{r}_B} |j\rangle, \quad (56)$$

For notational simplicity, we denote the internal states collectively by $j$, without specifying separately the states of $A$ and $B$.

The total density matrix associated with this outgoing wavefunction is written as:

$$\rho_{j,\vec{r}_A,\vec{r}_B,j',\vec{r}_A',\vec{r}_B'} = \int d\theta d\phi d\theta' d\phi' \sin\theta \sin\theta' \, \tilde{f}_{j_0 \to j}(\theta,\phi) \tilde{f}_{j_0 \to j'}^*(\theta',\phi')$$
$$e^{i(\frac{m_A}{M}\vec{K}+k_j\hat{k}(\theta,\phi))\vec{r}_A} e^{i(\frac{m_B}{M}\vec{K}-k_j\hat{k}(\theta,\phi))\vec{r}_B} \quad (57)$$
$$e^{-i(\frac{m_A}{M}\vec{K}+k_{j'}\hat{k}'(\theta',\phi'))\vec{r}_A'} e^{-i(\frac{m_B}{M}\vec{K}-k_{j'}\hat{k}'(\theta',\phi'))\vec{r}_B'}$$

Tracing out the external degrees of freedom $\vec{r}_A$ and $\vec{r}_B$ yields the reduced density matrix for the internal degrees of freedom:

$$\rho_{j,j'} = \int d\theta d\phi d\theta' d\phi' \sin\theta \sin\theta' S_j(\theta,\phi) S_{j'}^*(\theta',\phi')$$
$$\int d\vec{r}_A e^{i(k_j \hat{k}(\theta,\phi) - k_{j'} \hat{k}'(\theta',\phi'))\vec{r}_A} \qquad (58)$$
$$\int d\vec{r}_B e^{-i(k_j \hat{k}(\theta,\phi) - k_{j'} \hat{k}'(\theta',\phi'))\vec{r}_B}$$

The integrals over $\vec{r}_A$ and $\vec{r}_B$ impose delta-function constraints of the form $k_j \hat{k}(\theta,\phi) = k_{j'} \hat{k}'(\theta',\phi')$ whose validity depends on whether the internal energies associated with $|j\rangle$ and $|j'\rangle$ are degenerate.
If $k_j = k'_j$, the delta functions reduce to $\delta(\theta - \theta')\delta(\phi - \phi')$, leading to

$$\rho_{j,j'} = \int d\theta d\phi \sin\theta \tilde{f}_j(\theta,\phi) \tilde{f}_{j'}^*(\theta,\phi) \qquad (59)$$

On the other hands, if $k_j \neq k'_j$, the expression $k_j \hat{k}(\theta,\phi) = k_{j'} \hat{k}'(\theta',\phi')$ can never be respected and therefore, all coherences between different internal states vanish, and the reduced density matrix becomes diagonal:

$$\rho_{j,j'} = \int d\theta d\phi \sin\theta |\tilde{f}_j(\theta,\phi)|^2 \delta_{j,j'} \qquad (60)$$

From the reduced density matrix, the entanglement entropy is calculated with its eigenvalues $\lambda_i$

$$S_{ent} = -\sum_i \lambda_i \log_2 \lambda_i. \qquad (61)$$

## S6: Regularization procedure for the calculation of the DV-CV entanglement entropy

The modified scattering amplitude is defined as [1]:

$$\tilde{f}_{j_0 \to j}(\theta,\phi) = \delta(\Omega - \Omega_0)\delta_{j,j_0} + \frac{i}{2\pi}\sqrt{k_j k_0} f_{j_0 \to j}(\theta,\phi). \qquad (62)$$

In order to obtain finite values for entanglement measures, the angular delta function must be regularized. This regularization is implemented by replacing the delta distribution with a narrow, $L^2$-normalized Gaussian function $g(\theta,\phi)$. Physically, this corresponds to describing the incoming state as a narrow Gaussian wavepacket in angular space rather than an ideal plane wave. We choose

$$g(\theta,\phi) = \frac{1}{\sqrt{2\pi}\Sigma} exp\left[-\frac{\theta^2}{4\Sigma^2} - \frac{\phi^2}{4\Sigma^2}\right] \qquad (63)$$

where $\Sigma$ characterizes the angular width of the incoming wavepacket and is assumed to be small, $\Sigma << 1$. The function $g(\theta,\phi)$ is $L^2$ normalized:

$$\int d\theta d\phi \sin\theta |g(\theta,\phi)|^2 = 1 \qquad (64)$$

and satisfies a delta-function–like property under angular integration. Specifically, for any smooth angular function $h(\theta,\phi)$

$$\int d\theta d\phi \sin\theta g(\theta,\phi) h(\theta,\phi) = h(0,0) \qquad (65)$$

The diagonal elements of the reduced density matrix are obtained by integrating the squared modulus of the regularized generalized scattering amplitudes over the scattering angles:

$$\rho_{j,j} = \int d\theta d\phi \sin\theta |\tilde{f}_{j_0 \to j}(\theta,\phi)|^2. \qquad (66)$$

Substituting the regularized expression for $\tilde{f}_{j_0 \to j}(\theta, \phi)$ yields:

$$\rho_{j,j} = \int d\theta d\phi \sin\theta |g(\theta,\phi)|^2 \delta_{j,j_0} - \frac{k_0}{\pi} \int d\theta d\phi \sin\theta g(\theta,\phi) Im[f_{j_0 \to j_0}(\theta,\phi)]\delta_{j,j_0} + \frac{k_0 k_j}{4\pi^2} \int d\theta d\phi \sin\theta |f_{j_0 \to j}(\theta,\phi)|^2 \tag{67}$$

The first term evaluates to unity as a direct consequence of the $L^2$ normalization of the function $g(\theta, \phi)$. The second term simplifies by exploiting the delta-function–like property of the function $g(\theta, \phi)$ in the narrow-packet limit, together with the optical theorem, $Im[f_{j_0 \to j_0}(0,0)] = \frac{k_0}{4\pi}\sigma$, which gives :

$$-\frac{k_0}{\pi} \int d\theta d\phi \sin\theta g(\theta,\phi) Im[f_{j_0 \to j_0}(\theta,\phi)]\delta_{j,j_0} = -\frac{k_0^2}{4\pi^2}\sigma \delta_{j,j_0} \tag{68}$$

The third term is evaluated using the definition of the state-to-state integral cross section:

$$\frac{k_0 k_j}{4\pi^2} \int d\theta d\phi \sin\theta |f_{j_0 \to j}(\theta,\phi)|^2 = \frac{k_0^2}{4\pi^2}\sigma_{j_0 \to j} \tag{69}$$

Combining these contributions, we recover the expression for the diagonal elements of the reduced density matrix reported in the main text:

$$\rho_{jj} = (1 - \frac{k_0^2}{4\pi^2}\sigma)\delta_{j,j_0} + \frac{k_0^2}{4\pi^2}\sigma_{j_0 \to j}, \tag{70}$$

The off-diagonal element between degenerate final internal states $j$ and $j'$ are obtained from the angular overlap of the corresponding regularized generalized scattering amplitudes:

$$\rho_{j,j'} = \int d\theta d\phi \sin\theta \tilde{f}_{j_0 \to j}(\theta,\phi) \tilde{f}^*_{j_0 \to j'}(\theta,\phi). \tag{71}$$

Substituting the regularized expression for $\tilde{f}_{j_0 \to j}(\theta, \phi)$ yields:

$$\begin{aligned}
\rho_{j,j'} &= \int d\theta d\phi \sin\theta |g(\theta,\phi)|^2 \delta_{j_0 j}\delta_{j_0 j'} \\
&+ \frac{i\sqrt{k_j k_0}}{2\pi} \int d\theta d\phi \sin\theta f_{j_0 \to j}(\theta,\phi) g(\theta,\phi)\delta_{j_0 j'} \\
&- \frac{i\sqrt{k_{j'} k_0}}{2\pi} \int d\theta d\phi \sin\theta f^*_{j_0 \to j'}(\theta,\phi) g(\theta,\phi)\delta_{j_0 j} \\
&+ \frac{k_0 \sqrt{k_j k_{j'}}}{4\pi^2} \int d\theta d\phi \sin\theta f_{j_0 \to j}(\theta,\phi) f^*_{j_0 \to j'}(\theta,\phi)
\end{aligned} \tag{72}$$

The first term is simplified by the $L^2$ normalization of the function $g(\theta, \phi)$. The second and third terms simplify in the narrow-wavepacket limit, where $g(\theta, \phi)$ acts as an angular delta distribution. Collecting these contributions yields:

$$\rho_{j,j'} = \delta_{j_0 j}\delta_{j_0 j'} + \frac{i\sqrt{k_j k_0}}{2\pi} f_{j_0 \to j}(0,0)\delta_{j_0 j'} - \frac{i\sqrt{k_{j'} k_0}}{2\pi} f^*_{j_0 \to j'}(0,0)\delta_{j_0 j} + \frac{k_0 \sqrt{k_j k_{j'}}}{4\pi^2} \int d\theta d\phi \sin\theta f_{j_0 \to j}(\theta,\phi) f^*_{j_0 \to j'}(\theta,\phi) \tag{73}$$

Introducing the compact definitions

$$C_j = \frac{i}{2\pi}\sqrt{k_0 k_j} f^*_{j_0 \to j}(\Omega_0), \tag{74}$$

and

$$D_{j,j'} = \frac{k_0}{4\pi^2}\sqrt{k_j k_{j'}} \int d\theta d\phi \sin\theta f_{j_0 \to j}(\theta,\phi) f^*_{j_0 \to sj'}(\theta,\phi), \tag{75}$$

the off-diagonal elements of the reduced density matrix can be written in the compact form reported in the main text:

$$\rho_{j,j'} = \delta_{jj_0}\delta_{j'j_0} + \delta_{jj_0}C^*_{j'} + \delta_{j'j_0}C_j + D_{j,j'}, \tag{76}$$

# S7: DV-DV entanglement entropy for ultracold Rb + SrF collisions: full results

Calculations for the DV-DV etanglement for ultracold Rb + SrF collisions were performed for several initial internal states, and the complete set of results is provided in the following table.

| Initial state | $S_{ent}^{int-int}$ |
|---|---|
| $\|2\rangle_{SrF}\|1\rangle_{Rb}$ | 0.74 |
| $\|2\rangle_{SrF}\|2\rangle_{Rb}$ | 0.69 |
| $\|2\rangle_{SrF}\|3\rangle_{Rb}$ | 0 |
| $\|3\rangle_{SrF}\|1\rangle_{Rb}$ | 0 |
| $\|3\rangle_{SrF}\|2\rangle_{Rb}$ | 0.41 |
| $\|3\rangle_{SrF}\|3\rangle_{Rb}$ | 0.38 |
| $\|4\rangle_{SrF}\|1\rangle_{Rb}$ | 0 |
| $\|4\rangle_{SrF}\|2\rangle_{Rb}$ | 0.81 |
| $\|4\rangle_{SrF}\|3\rangle_{Rb}$ | 0.84 |

# S8: Details on the S-matrix scattering calculations

For ultracold Rb + SrF collisions, rigorous coupled-channel (CC) calculations were performed at a collision energy of 100 nK. The CC equations were integrated numerically using the log-derivative propagator, yielding the asymptotic radial solutions from which the S-matrix elements and state-to-state scattering cross sections were extracted as a function of magnetic field, following the procedure described in Ref. [3]. The basis set included the two lowest total rotational angular momentum blocks ($J_r^{\max} = 1$) and a rotational basis for SrF extending up to $N_{\max} = 175$, ensuring convergence of the scattering observables.

For the Rb + Sr$^+$ system, we performed CC calculations based on state-of-the-art *ab initio* interaction potentials and second-order spin–orbit couplings, as detailed in Ref. [4, 5]. Numerical convergence was verified using extended basis sets incorporating partial waves up to $\ell = 2$, which is essential due to the long-range atom–ion interaction scaling as $1/R^4$ and the resulting contributions from *s*-, *p*-, and *d*-wave channels.

The S-matrix elements for the F + HD chemical reaction were computed by solving the coupled-channel (CC) scattering equations in hyperspherical coordinates using the ABC code [6], employing the Stark–Werner potential energy surface (PES) [7]. Calculations were performed at an incident collision energy of 1 K, including total angular momentum values $J = 0$–7 and inversion parities $\epsilon = +1$ for $J = 0$ and $\epsilon = \pm 1$ for $J > 0$. The CC equations were propagated up to a maximum hyperradius of $\rho_{\max} = 40$ a.u. with a step size of $\Delta \rho = 0.01$ a.u. A cutoff energy of 2.5 eV was imposed for the rovibrational basis set, allowing for up to 15 rotational levels in each reaction arrangement. The maximal helicity quantum number was set to $k_{\max} = 4$ for $J < 5$ and to $k_{\max} = J$ for $J \geq 5$, ensuring numerical convergence. The S-matrix elements were initially obtained in the body-fixed frame (BFF) and subsequently transformed to the space-fixed frame (SFF) [8]. The transformation matrix between the two frames was constructed by diagonalizing the operator $\hat{L}^2$, where $\vec{L}$ denotes the orbital angular momentum of the three-body collision complex. The resulting S-matrix elements and state-resolved reaction probabilities were converged to within $\simeq 10\%$.